\documentclass[twocolumn,superscriptaddress,showpacs]{revtex4-1}
\usepackage{amsmath}
\usepackage{latexsym}
\usepackage{amssymb}
\usepackage{mathtools}
\usepackage{braket}
\usepackage{graphicx}
\usepackage{wrapfig}
\usepackage{dsfont}
\usepackage{etoolbox}
\usepackage[colorlinks=true, citecolor=blue, urlcolor=blue]{hyperref}
\usepackage{float}
\usepackage{amsfonts}
\usepackage{makecell}

\begin{document}

\title{Measurement Based Quantum Heat Engine with Coupled Working Medium}

\author{Arpan Das}
\email{arpandas@imsc.res.in}
\affiliation{Optics and Quantum Information Group, The Institute of Mathematical Sciences, CIT Campus, Taramani, Chennai 600113, India.}
\affiliation{Institute of Physics, Sachivalaya Marg,
Bhubaneswar 751005, Odisha, India.}
\affiliation{Homi Bhabha National Institute, Training School Complex, Anushakti Nagar, Mumbai 400085, India.}

\author{Sibasish Ghosh}
\email{sibasish@imsc.res.in}
\affiliation{Optics and Quantum Information Group, The Institute of Mathematical Sciences, CIT Campus, Taramani, Chennai 600113, India.}
\affiliation{Homi Bhabha National Institute, Training School Complex, Anushakti Nagar, Mumbai 400085, India.}

\begin{abstract}
We consider measurement based single temperature quantum heat engine without feedback control, introduced recently by {Yi, Talkner and Kim} [{\it Phys. Rev. E}  {\bf 96}, 022108 (2017)].~Taking~the working medium of the engine to be a one-dimensional Heisenberg model of two spins, we calculate the efficiency of the engine undergoing a cyclic process. Starting with two spin-1/2 particles, we~investigate the scenario of higher spins also. We show that, for this model of coupled working medium, efficiency can be higher than that of an uncoupled one. However, the  relationship  between the coupling constant and the efficiency of the engine is rather involved. We find that in the higher spin scenario efficiency can sometimes be negative (this means work has to be done to run the engine cycle) for certain range of coupling constants, in contrast to the aforesaid work of { Yi, Talkner and Kim}, where~they showed that the extracted work is always positive in the absence of coupling. We provide arguments for this negative efficiency in higher spin scenarios. Interestingly, this happens only in the asymmetric scenarios, where the two spins are different. Given these facts, for judiciously chosen conditions, an  engine with coupled working medium gives advantage for the efficiency over the uncoupled one.
\end{abstract}

% Keywords
%\keyword{quantum heat engine; measurement driven engine}

\maketitle
%\begin{document}

\section{Introduction}
Unification of seemingly different heat engines in terms of efficiency started in the early 19th century with Sadi Carnot~\cite{carnot}. From~then onward, converting heat into useful work with increasing efficiency got a thrust in practical and industrial territories. For~a standard heat engine, working cyclically between two heat baths of temperature $T_1$ and $T_2$ ($T_2<T_1$), efficiency of the engine is upper bounded by \mbox{$\eta=1-T_2/T_1$}, the~Carnot efficiency~\cite{callen}. The second law of thermodynamics puts this fundamental limitation on the extent of work that can be converted from heat.~The~laws of thermodynamics are empirical and were first adopted for classical macroscopic systems.~Naturally, the~validity of the laws of thermodynamics are questionable and subject to verification in the quantum regime. Moreover, quantum mechanics gives the dynamical viewpoint of thermodynamics~\cite{kosloff, kosloff2}, describing~the emergence of thermodynamic laws from quantum~mechanics.
 
The idea of quantum heat engine   first appeared in a paper by Scovil and Schulz-DuBois~\cite{scovil}, where the authors demonstrated that three level  masers can be treated as a working medium for heat engines. Today, the  study of heat engines in quantum domain is an active area of research both due to the gradual miniaturization of current technology as well as its theoretical~richness. 

Within the quantum engines scenario, quantum analog of the classical heat engines~\cite{kieu, nori} and many other generalizations~\cite{alicki, kosloff3, scully, levy} have been studied. Analysis of finite power quantum heat engines also have a significant amount of literature (e.g.,  ~\cite{geva,  feldman, aleya, esposito, campisi, sir, holubec}).~With the onset of the quantum effects, many interesting phenomena  such as the increase of efficiency beyond Carnot's limit~\cite{gs, lutz, lutz1} may occur. However,~it is not in contradiction with the second law of thermodynamics. The~compensation comes from carefully accounting all the work costs.~ \citet{deffner} showed that if one accounts for the work cost to maintain the non-equilibrium reservoir, Carnot's limit cannot be surpassed. However, understanding quantum thermodynamic machines~\cite{arnab} and the role of quantum effects~\cite{huber, huber2, alickientan, coherence, huber1, chiara} in quantum thermodynamics is far from fully understood. Where~quantum effects set a limit to our ability~\cite{horodecki} for practical purposes and where we can actually use the quantum resources are still parts of ongoing research field. The~approach to resource theory of quantum thermodynamics~\cite{nelly} tells us about the fundamental corrections to the laws of thermodynamics, setting~the limit to the performance of quantum heat~engines. 

Previously, it was shown that~\cite{george, ferdi} a quantum Otto engine with coupled working medium leads to a higher efficiency than that of an uncoupled one. In addition, in~information heat engine, e.g., the Szilard engine, where exploiting the information one can extract work from an engine operating at a single temperature~\cite{szilard, maxwell, kim, kim1}, it was shown that~\cite{aleya1, zurek1, lidia, watanbe} entanglement can be used to extract work beyond the limit, which is possible using classical correlation~only. 

Recently, in ~\cite{talkner1}, the~authors   introduced a new kind of single temperature quantum heat engine without feedback control. The~essential part of the engine which replaces feedback is a non-selective quantum measurement on the working medium, changing the average energy of the system, and,~thus, enabling one to extract useful work. This engine is  similar to a quantum Otto cycle~\cite{kosloff_otto, suman1, suman2, brown} with one thermalization stroke being replaced by a quantum non-selective measurement, whereas, in  Maxwell's demon and Szilard engine~\cite{szilard, maxwell}, work is extracted from a single heat reservoir using feedback control.  Another version of Maxwell's demon engine was introduced in~\cite{elouard1, elouard2}, where without the presence of any thermal bath, work can be extracted using measurement and feedback control. Thus, quantum measurement plays an important role in quantum thermodynamics. Energetic cost for performing a measurement~\cite{faist, reeb, elouard3, janet, faist2} and using the average energy change due to the measurement for extracting useful work are two important facets of quantum thermodynamics.
In a subsequent work~\cite{talkner2}, the~authors calculated the detailed fluctuation of work and heat in the above said measurement driven single temperature heat engine without feedback control as well as   considered the finite power~scenario. 

In this paper, we analyze the role of coupled working medium in this single temperature measurement driven quantum heat engine without feedback control~\cite{talkner1}.~Taking a coupled one-dimensional Heisenberg model as the working medium, we show an advantage for the efficiency over the uncoupled~one. 

First, we start with a one-dimensional Heisenberg model of two spin-1/2 particles and then generalize that for two spin-$d$ particles, where   $d$ can take values of  $1/2$, $1$, and~$3/2$. We note that, for different choices of non-selective measurements, the~efficiency of the heat engine changes. In addition, in the higher-dimensional scenario, another interesting feature is observed: we can either extract work from the engine cycle or have to invest work to run the cycle depending upon the spin configuration we choose. By~judiciously choosing all the conditions, such as  measurement choices, coupling constant and the dimension of the Hilbert spaces, one can optimize the engine performance, which is better than the uncoupled one in terms of~efficiency.

The paper is organized as follows. In~the next section, we give a short introduction about the single temperature measurement-based heat engine {as introduced in \cite{talkner1}}. In~Sections~\ref{sec4} and \ref{sec5}, we present our result for the {coupled measurement based heat engine} taking the working medium to be the Heisenberg model of two spin half particles. In~Section~\ref{sec6}, we consider the higher-dimensional scenario. In~  Section~\ref{sec7}, we present an analysis of the global and local work. Finally, we conclude in Section~\ref{conclu}.

\section{Single Temperature Measurement Driven Quantum Heat Engine without~Feedback}
\label{sec3}
In this section, we briefly discuss the recently introduced  measurement based single temperature quantum heat engine without feedback control~\cite{talkner1}. It is very similar to the Otto cycle except for one thermalization step, which is replaced by a non-selective quantum measurement. If~it had been a classical system, in~principle, there would be no subsequent effect of the measurement on the system. However,~the quantum mechanical system is generally disturbed by measurement and hence average energy of the system changes. Judiciously choosing the measurement operators, as discussed in~\cite{talkner1}, we can extract work form this type of engine. We now briefly describe the engine~cycle.

The working medium of the engine has a Hamiltonian $H(\lambda)$, which is a function of an external control parameter $\lambda$.~The~system starts from a thermal state of temperature $T$. This can be achieved with the help of a heat bath of temperature $T$, which is the only heat bath to be used throughout the action of the engine. The~system is brought to the  contact with the bath and   the system allowed to thermalize. After~a long enough time, when the system attains equilibrium thermal state, the~heat bath is detached and it gets ready for the first cycle of our heat engine. Thus, the~initial state of the system is  $\rho_{int}=e^{-\beta H(\lambda_{int})}/Z=\sum_n (e^{-\beta E_n(\lambda_{int})}/Z)\ket{n(\lambda_{int})}\bra{n(\lambda_{int})}$, where       $Z=\sum_n e^{-\beta E_n(\lambda_{int})}$, while  $\ket{n(\lambda)}$ and $E_n(\lambda)$ are, respectively, the $n$th eigenstate and eigenvalue of the Hamiltonian $H(\lambda)$. Now, the~engine strokes are as~following.

{\bf{\it First stroke}}: The first stroke of the cycle is an adiabatic compression process.~The~working system is isolated from the heat bath and the Hamiltonian is changed quasi-statically from $H(\lambda_{int})$ to $H(\lambda_{fin})$, with~initial occupation probabilities of the state remaining unchanged. For~a system defined by a density matrix $\rho$ and Hamiltonian $H$, its internal energy or average energy is defined as $U=Tr[\rho H]$. Change in internal energy is the sum of two contributions~\cite{sai}, one is heat, defined~as $dQ=Tr[Hd\rho]$, and the other is work, defined as $dW=Tr[\rho dH]$.~However,    this identification of heat and work is not always valid, especially in strong system bath coupling~\cite{jar, lobet}. We   follow here the above definition. {Although, during a general adiabatic process, the~state of the working medium changes,   for~the model of engine cycle  we consider in this paper, the~state of the working medium does not change throughout the adiabatic stroke.} Hence, in~the first stroke, change in the internal energy of the system is  $W_1=Tr[\rho_{int}(H(\lambda_{fin})-H(\lambda_{int}))]$, which can also be written as  $W_1=\sum_n [E_n(\lambda_{fin})-E_n(\lambda_{int})]p_n (\lambda_{int})$, where $p_n (\lambda_{int})=(e^{-\beta E_n(\lambda_{int})}/Z)$. If~this is positive, then this is the energy gained by the system. Thus, the~average work that is extracted from this stroke is $-W_1$.%italic is just to highlight four different strokes of the engine. It is not necessary. 

{\it Second stroke}: The next stroke is the most crucial and special one, which involves a non-selective measurement. A~measurement~\cite{nielson} corresponding to an observable $\hat{G}$ can be described by a POVM, $\{G_n\}$, where       $G_n\geq 0$ are the POVM effects, $\sum_n G_n=\mathds{1}$ and $Tr[\rho G_n]$ is the probability of getting $n$th outcome denoted here as $\alpha_n$. If~in addition the POVM elements satisfy $G_mG_n=\delta_{mn}G_n$, then~they are orthogonal projectors and the observable $\hat{G}$ can be written as $\hat{G}=\sum_n \alpha_n G_n$ (spectral value decomposition), where $\alpha_n$s are now the eigenvalues of $\hat{G}$. Equivalently, measurement can be completely described by a set of measurement operators $\{M_n\}$, with~${{M}_n^{\dagger}} M_n=G_n$. If~the state of the system before the measurement is $\ket{\phi}$, then corresponding to the $n$th outcome, the~state of the system after measurement will be  $\frac{M_n\ket{\phi}}{\sqrt{p_n}}$ or $\frac{M_n\rho_{int}{M}_n^{\dagger}}{p_n}$, where $\rho_{int}=\ket{\phi}\bra{\phi}$ and $p_n=Tr[\rho_{int}{{M}_n^{\dagger}} M_n]=\bra{\phi}{{M}_n^{\dagger}} M_n\ket{\phi}$ is the probability of getting $n$th outcome. In~a non-selective measurement, i.e.,~if the outcomes of the measurements are not recorded, then the state after measurement is $\sum_n M_n\rho_{int}{M}_n^{\dagger}$.  Thus, after~the first stroke, we do a non-selective measurement described by the measurement operators $\{M_\alpha\}$ on the system state giving the post-measurement state as $\rho_M=\sum_\alpha M_\alpha\rho_{int}{M}_\alpha^{\dagger}$.
In this stroke, the~Hamiltonian of the system is unchanged at $H(\lambda_{fin})$. Thus, the~average energy change of the system is given by  $Q_M=Tr[(\rho_M-\rho_{int})H(\lambda_{fin})]$, which is   reminiscent of heat. This can be written as~\cite{talkner1},
\begin{eqnarray}
%\nonumber
&&Q_M=\sum_{m,n}[E_m(\lambda_{fin})-E_n(\lambda_{fin})]T_{m,n}p_n(\lambda_{int})\\
\nonumber
%&&=\frac{1}{2}\sum_{m,n}[E_m(\lambda_{fin})-E_n(\lambda_{fin})]T_{m,n}[p_n(\lambda_{int})-p_n(\lambda_{fin})]
&&=\sum_n\bra{n(\lambda_{fin})}H_M(\lambda_{fin})-H(\lambda_{int})\ket{n(\lambda_{fin})}p_n(\lambda_{int}).
\end{eqnarray}
where       $T_{m,n}=\sum_\alpha|\bra{n(\lambda_{fin})}M_\alpha \ket{m(\lambda_{fin})}|^2$   is the transition probability from a eigenstate labeled $n$ before the measurement to an eigenstate labeled $m$ after the measurement, and $H_M(\lambda_{fin})=\sum_\alpha M_\alpha H(\lambda_{fin})M_\alpha$. As~shown in~\cite{talkner1}, $Q_M$ is always positive, which follows from the properties of the transition matrix, $T_{m,n}=T_{n,m}$ and $\sum_n T_{m,n}=1$.  This fact   is illustrated more explicitly in the next section.
It is also noted that, whenever the Hamiltonian $H(\lambda_{fin})$ of the system does not commute with the measurement operators, we get a nonzero $Q_M$.%changed . to , please confirm.---Yes, it was a typo.

{\it  Third stroke}: This is the second adiabatic process. The~Hamiltonian $H(\lambda_{fin})$ is very slowly changed back to the initial Hamiltonian $H(\lambda_{int})$, with~unchanged occupation probabilities of the state. As in the previous adiabatic stroke, the~average change in energy of the system is  $W_2=Tr[\rho_{M}(H(\lambda_{int})-H(\lambda_{fin}))]$, { as the state of the system is unchanged throughout the stroke. This} is nothing but $\sum_n [E_n(\lambda_{int})-E_n(\lambda_{fin})]p_n^M $, where       $p_n^M=\bra{n(\lambda_{fin})}\rho_M \ket{n(\lambda_{fin})}=\sum_m p_m(\lambda_{int})T_{m,n}$ is the probability of finding the $n$th eigenstate of $H(\lambda_{fin})$ in $\rho_M$. Thus, the~work extracted form this adiabatic stroke is $-W_2$,  which follows from the similar argument given in the description of the first~stroke. 

{\it Fourth stroke}: In this last stroke of the cycle, the~system is brought into contact with the heat bath of temperature $T$, while keeping the Hamiltonian fixed at $H(\lambda_{int})$ and allowed to thermalize, until~it goes back to the initial thermal state $\rho_{int}$. Thus, heat transfer for this stroke is given by 
$Q_T=Tr[(\rho_{int}-\rho_{M})H(\lambda_{int})]$, {as the Hamiltonian is fixed in this stroke.~This} can be written as  $Q_T=\sum_n E_n(\lambda_{int})[p_n(\lambda_{int})-p_n^M]$ and is shown to be negative in~\cite{talkner1}.~This means that heat is going to the heat bath from the working medium at this~stage. 

The signs of each of these quantities, e.g.,~$W_1$, $Q_M$, $W_2$, and~$Q_T$, are  analyzed explicitly in the next section in the case of coupled working medium. Thus, the~whole cycle is, similar to energy $Q_M$,   taken by the system, doing a work $-(W_1+W_2)$ and dumping energy $Q_T$ to a heat bath. Thus, we have $Q_M+Q_T=-(W_1+W_2)$, correctly depicting the first law of thermodynamics, i.e energy conservation. Efficiency of the heat engine is given { as the ratio of extracted work $-(W_1+W_2)$ over the average energy change $Q_M$ in the measurement stroke}: $\eta=\frac{-(W_1+W_2)}{Q_M}$.

 \citet{talkner1} showed that {the extracted work} $-(W_1+W_2)$ is always positive, relying~on the fact that work strokes are either adiabatic expansion or compression of the working medium. However, ~we   show in the next section that, for a coupled working medium, their way of reasoning does not hold well and there can be instances where $-(W_1+W_2)$ is~negative.

\section{Coupled Single Temperature Measurement~Engine}
\label{sec4}
In this section, we present an analysis of a coupled measurement-based single temperature heat engine. We consider the working medium of the system to be a one-dimensional Heisenberg model of two particles with the following Hamiltonian:
\begin{equation}
\label{hamilstruc}
H=8J\vec{S_A}.\vec{S_B}+2B(S_A^z+S_B^z).
\end{equation}

For two spin half particles, $\vec{S_A}=\vec{S_B}=\frac{1}{2}\vec{\sigma}$, where $\vec{\sigma}=(\sigma_x,\sigma_y,\sigma_z)$ are the Pauli matrices.
Thus, in~this case, we can write the Hamiltonian as,
\begin{eqnarray}
\nonumber
&H=2J({\sigma_x}^A\otimes{\sigma_x}^B+{\sigma_y}^A\otimes{\sigma_y}^B+{\sigma_z}^A\otimes{\sigma_z}^B)\\
&+B({\sigma_z}^A\otimes \mathds{1}^B+\mathds{1}^A\otimes{\sigma_z}^B),
\end{eqnarray}
where       $J$ is the coupling constant and $B$ is the external magnetic field.~The~entanglement between two qubits for this model has been studied in~\cite{vedral}.~$J>0$ and $J<0$ cases correspond to the anti-ferromagnetic and ferromagnetic interactions, respectively. In~this paper, we restrict ourselves to the anti-ferromagnetic case only.
Eigenvalues and eigenstates of this Hamiltonian are listed in Table~\ref{tabtwohalf}, where       $\ket{0}\doteq
\begin{pmatrix} 1\\0\end{pmatrix}$ and $\ket{1}\doteq
\begin{pmatrix} 0\\1\end{pmatrix}$ are the eigenstates of $\sigma_z$. As~described above, the engine~cycle of the measurement based heat engine has four steps.
The first stroke of the cycle is an adiabatic compression, where the Hamiltonian of the working system, as~described above, is quasi-statically changed from an initial parameter value to a final parameter value. External magnetic field $B$ is the parameter of the Hamiltonian here. It is changed quasi-statically from the initial value $B_1$ to the final value $B_2$. As~this process is done adiabatically, the~state of the system has same occupation probabilities throughout the stroke. At~the beginning of the cycle, we take the working medium of the heat engine to be in a thermal equilibrium state of temperature $T$. Thus, here, $\rho_{int}=\sum_{n=1}^4 P_n \ket{\psi_n(B)}\bra{\psi_n(B)}$, where       $P_n=\exp (-E_n/k_BT)/Z$, $Z=\sum_n \exp (-E_n/k_BT)$, $E_n'$s and corresponding $\ket{\psi_n(B)}$ are given in Table~\ref{tabtwohalf}. Then, in the second stroke of the cycle, the~Hamiltonian of the system is kept unaltered but a measurement of an observable is performed on the system. As    discussed  above, the~observable has to be non-commuting with the Hamiltonian to get a positive work output. In~this case, we have a distributed system and we   show  that the efficiency of the heat engine depends on the local measurements we are performing. The third stroke is again an adiabatic process changing the external magnetic field $B_2$ back to $B_1$.
The final stage of the cycle is a thermalization step and in this stage the system is brought to contact with a heat bath of the starting temperature $T$ and   the system allowed to thermalize for a sufficiently long time, after~which it again goes back to the initial thermal equilibrium state.
Now, as~mentioned above, the~initial state of the system is a thermal state of temperature $T$,
\begin{equation}
\rho_{int}=\sum_{n=1}^4 P_n \ket{\psi_n(B)}\bra{\psi_n(B)},
\label{initialthermal}
\end{equation}
where       $P_n=\exp (E_n/k_BT)/Z$, $Z=\sum_n \exp (E_n/k_BT)$, $E_n$ and $\ket{\psi_n(B)}$s are the energy eigenvalues and eigenstates, respectively, as listed in     Table~\ref{tabtwohalf}. 
From now on, we     take $k_BT=1$ throughout the paper. The~energy eigenvalues and hence the probabilities $P_n$ depend on the changing parameter, which is the external magnetic field $B$, but~the eigenstates are independent of the parameter. Thus, we   omit the dependence of $B$ from the notations of the eigenstates. 
The average work in the first adiabatic stroke is,
\begin{equation}
W_1=\sum_n[E_n(B_2)-E_n(B_1)]P_n(B_1),
\label{firstwork}
\end{equation}
as the state remains in its instantaneous eigenstate with same probability. For~the system we consider  and the initial thermal state of the system, we have 
\begin{equation}
W_1=\frac{2(B_1-B_2)(-1+e^{4B_1})}{1+e^{2B_1}(1+e^{2B_1}+e^{8J})}.
\label{work1}
\end{equation}
\vspace{-12pt}
\begin{table}[h]
\centering
\caption{$S_A=1/2$,$S_B=1/2$.}\label{tabtwohalf}
\begin{tabular}{cccc}
%\hline
%Acetylene: hydrogen concentration & \multicolumn{3}{|r|}{Average roughness values $S_a$ in nm} \\[0.5 ex]
%\hline
%{} &{|r|}{Value of $p$ to start violation} \\[0.3 ex]
%\toprule
\hline
%{} & {}\\
%{} &\multicolumn{2}{|r|}{Value of $p$ to start violation} \\[0.3 ex]
%\hline
%\cline{1-2}
{\textbf{Eigenvalues}}  &   \textbf{Eigenstates} \\
%\midrule
\hline
$2J+2B=E_4$ &$\ket{00}=\ket{\psi_4(B)}$\\

$2J=E_3$ & $\sqrt{\frac{1}{2}}(\ket{10}+\ket{01})=\ket{\psi_3(B)}$\\

$2J-2B=E_2$ & $\ket{11}=\ket{\psi_2(B)}$\\

$-6J=E_1$ & $\sqrt{\frac{1}{2}}(\ket{10}-\ket{01})=\ket{\psi_1(B)}$\\
%\bottomrule
\hline
\end{tabular}
\end{table}

The next stroke is the non-selective measurement part. We can choose any arbitrary observable for measurement with only constraint that the observable must be non-commuting with the Hamiltonian. For~the time being, we restrict ourselves for projective measurements.~In~our case, for~coupled measurement based heat engine, we take most general measurement operators as,
\begin{eqnarray}
&&M_1=\ket{+^a}\bra{+^a}\otimes \ket{+^b}\bra{+^b} \\
&&M_2=\ket{+^a}\bra{+^a}\otimes \ket{-^b}\bra{-^b}\\
&&M_3=\ket{-^a}\bra{-^a}\otimes \ket{+^b}\bra{+^b}\\
&&M_4=\ket{-^a}\bra{-^a}\otimes \ket{-^b}\bra{-^b},
\end{eqnarray}
where       $\ket{+^a}\bra{+^a}$ and $\ket{-^a}\bra{-^a}$ are the eigenstate projectors for the observable $\vec{\sigma}.\hat{a}$ for one party and  $\ket{+^b}\bra{+^b}$ and $\ket{-^b}\bra{-^b}$ are the eigenstate projectors for the observable $\vec{\sigma}.\hat{b}$ for the other. 
%These measurement operators can also be written as,
%\begin{eqnarray}
%&M_1=\frac{1}{4}(\mathds{1}+\vec{\sigma}.\hat{n})\otimes (\mathds{1}+\vec{\sigma}.\hat{m})\\
%&M_2=\frac{1}{4}(\mathds{1}+\vec{\sigma}.\hat{n})\otimes (\mathds{1}-\vec{\sigma}.\hat{m})\\
%&M_3=\frac{1}{4}(\mathds{1}-\vec{\sigma}.\hat{n})\otimes (\mathds{1}+\vec{\sigma}.\hat{m})\\
%&M_4=\frac{1}{4}(\mathds{1}-\vec{\sigma}.\hat{n})\otimes (\mathds{1}-\vec{\sigma}.\hat{m})
%\end{eqnarray}
Now, if the initial state of the working medium is given by Equation~(\ref{initialthermal}), then, for a non-selective measurement given by the above measurement operators, the~post measurement state will be,
\begin{equation}
\rho_{M}=\sum_{k=1}^4 M_k\rho_{int}M_k.
\end{equation}

The average energy change of the system during this measurement stroke is given by,
\begin{equation}
Q_M=\sum_{m,n}[E_n(B_2)-E_m(B_1)]T_{m,n}P_m(B_1),
\label{heatfirst}
\end{equation}
where       $T_{m,n}$ is the transition probability, written as 
\begin{equation}
T_{m,n}=\sum_k|\bra{\psi_m}M_k\ket{\psi_n}|^2\hspace{2mm} \forall m,n \in \{1,2,3,4\}.
\label{transition}
\end{equation}
%$\ket{\psi_m}$ and $\ket{\psi_n}$ denote the eigenstates of the Hamiltonian as written in     Table~\ref{tabtwohalf} and $M_k$s are the measurement operators written above.

For the most general form of the measurement operators, the~expressions are quite complicated. Thus, we first write the expressions for some special choices of measurements. The first case is when $\vec{\sigma}.\hat{a}$ is $\sigma_x$ and $\vec{\sigma}.\hat{b}$ is $\sigma_z$. For~this case,
\begin{equation}
Q_M=\frac{B_2(-1+e^{4B_1})-2[1+e^{2B_1}(1+e^{2B_1}-3e^{8J})]J}{1+e^{2B_1}(1+e^{2B_1}+e^{8J})}.
\label{heat2}
\end{equation}

For $\vec{\sigma}.\hat{a}=\sigma_y$ and $\vec{\sigma}.\hat{b}=\sigma_z$, the~expression for the average change in energy remains same. Another~case is when $\vec{\sigma}.\hat{a}=\sigma_x$ and $\vec{\sigma}.\hat{b}=\sigma_y$. In~this scenario,
\begin{equation}
Q_M=\frac{2B_2(-1+e^{4B_1})-2[1+e^{2B_1}(1+e^{2B_1}-3e^{8J})]J}{1+e^{2B_1}(1+e^{2B_1}+e^{8J})}.
\label{heat3}
\end{equation}

There are many other different choices of measurement operators, such as both   measuring $\sigma_x$ or $\sigma_y$ or $\sigma_z$, etc. In~each case, the~expression for $Q_M$ will change accordingly.
This average energy change is similar to  heat in the conventional quantum heat engine. Now, the~third step of the cycle is again an adiabatic process, where the magnetic field $B_2$ is changed back into $B_1$ very slowly. For~this part of the cycle, the~work done is,
\begin{equation}
W_2=\sum_n[E_n(B_1)-E_n(B_2)]P'_n,
\label{work2}
\end{equation}
where       $P'_n=\bra{\psi_n}\rho_{PM}\ket{\psi_n}=\sum_m T_{m,n}P_m(B_1)$ is the probability of getting $n$th eigenstate in the post measurement state. When  $\vec{\sigma}.\hat{a}=\sigma_x$ and $\vec{\sigma}.\hat{b}=\sigma_z$, the~expression of this work for our system is,
\begin{equation}
W_2=\frac{(B_1-B_2)(1-e^{4B_1})}{1+e^{2B_1}(1+e^{2B_1}+e^{8J})}.
\end{equation}

The last step of the cycle is to bring the system in contact with the heat bath of temperature $T$ and let it thermalize back to the initial thermal state $\rho_{int}$. Heat exchanged in this step is given by,
\begin{equation}
\label{heatcold}
Q_T=\sum_n E_n(B_1)(P_n(B_1)-P'_n).
\end{equation}

Again, for  $\vec{\sigma}.\hat{a}=\sigma_x$ and $\vec{\sigma}.\hat{b}=\sigma_z$, heat dumped into the heat bath in this last step is,
\begin{equation}
Q_T=-6J+\frac{B_1(1-e^{4B_1})+8(1+e^{2B_1}+e^{4B_1})J}{1+e^{2B_1}(1+e^{2B_1}+e^{8J})}.
\end{equation}

Thus, as~discussed in the previous section, the~total work that can be obtained from the cycle is given by the sum of the works that can be extracted in the first and third strokes. For~$\vec{\sigma}.\hat{a}=\sigma_x$ and $\vec{\sigma}.\hat{b}=\sigma_z$, { total extracted work is given by},
\begin{equation}
W_{t}=-W=-(W_1+W_2)=\frac{(B_1-B_2)(1-e^{4B_1})}{1+e^{2B_1}(1+e^{2B_1}+e^{8J})}.
\label{wglosphalf}
\end{equation}

The quantities we have calculated so far are global (from the perspective of two systems together), i.e.,~global heat, global work or global energy change. In~the next section, we    discuss the global efficiency of the heat engine.~In addition, the~measurements we choose are all projective measurements. We~investigated some cases of POVM, namely the SIC POVM~\cite{povm}, and~some other examples. However,~in all   cases, we found that projective measurements are more effective so far as the efficiency is concerned. Thus, we   be restricting ourselves with projective measurements~only.

\section{Efficiency of the Heat Engine, Global~Analysis}
\label{sec5}
In this section, we evaluate the efficiency of the measurement based coupled heat engine and compare it with the uncoupled one. Before~that, it is necessary to determine the signs of the quantities  $Q_M$ (Equation (\ref{heatfirst})), $Q_T$ (Equation (\ref{heatcold})), $W_1$ (Equation (\ref{firstwork})) and $W_2$ (Equation (\ref{work2})) for this coupled working medium. The~average energy change $Q_M$ of the system during the measurement stroke, given~in Equation~(\ref{heatfirst}), can also be written as~\cite{talkner1},
\begin{equation}
Q_M=\frac{1}{2}\sum_{m,n}[E_n(B_2)-E_m(B_1)]T_{m,n}(P_m(B_1)-P_n(B_1)).
%\label{heat1}
\end{equation}

This is obtained by employing the properties of the transition matrix, $T_{m,n}=T_{n,m}$ and $\sum_{m} T_{m,n}=1$. From~the fact that $T_{m,n}\geq 0$ and the equilibrium occupation probability $P_n$ for an energy level $E_n$ decreases with the increase of energy $E_n$, it turns out that $Q_M\geq 0$. That means, in~the measurement step, heat enters into the working medium. Similarly, the~expression for $Q_T$ given in Equation~(\ref{heatcold}) can be written as,
\begin{equation}
\label{heatcold1}
Q_T=\frac{1}{2}\sum_{n,m} (E_n(B_1)-E_m(B_1))T_{m,n}(P_n(B_1)-P_m(B_1)).
\end{equation}

By   similar arguments made for $Q_M$, it is evident that $Q_T\leq 0$, which means that  heat goes from the working medium to the heat bath of temperature $T$ at the last step of the cycle. Now, to determine the signs of $W_1$ and $W_2$, let us first write down the alternative expression for the work   as  done   in  ~\cite{talkner1}.
\begin{equation}
\label{negwork1}
W=-W_t=\frac{1}{2}\sum_{n,m}(\Delta^f_{m,n}-\Delta^i_{m,n})T_{m,n}[P_m(B_1)-P_n(B_1)],
\end{equation}
where       $\Delta^\alpha_{m,n}$ denotes the difference between the $m$th and $n$th energy eigenvalues and is given by,
$\Delta^\alpha_{m,n}\equiv E_m(\lambda_\alpha)-E_n(\lambda_\alpha)$ , for~$\alpha=i,f$, $\lambda_i=B_1$ and $\lambda_f=B_2$.

In   \cite{talkner1}, the authors designed the adiabatic strokes to be compression and expansion, in~the sense that spacing between the energy levels of the Hamiltonian either increases or decreases. Then,~they argued that for the compression stroke, $\Delta^f_{m,n}\geq \Delta^i_{m,n}$, and as the canonical probability decreases monotonically with the increase of energy, for~$\Delta^i_{m,n}>0$, $P_m(B_1)-P_n(B_1)$ is negative. Together with the fact that $T_{m,n}$ is positive, every term in the above expression of $W$ is non-positive and hence the total work extracted $W_t=-W$ is always positive. Similar arguments hold for the expansion stroke. However,~for a coupled working medium, this argument does not hold good. In~the presence of coupling $J$, the~uniform increase or decrease of spacing between the energy levels does not happen. For~example, for~the two spin-1/2 scenario, the~energy eigenvalues are $-6J$, $2J-2B$, $2J$, $2J+2B$, which~are ordered from low to high for $J>0$ (and for $4J\geq B$).~Throughout the paper, we   consider  the anti-ferromagnetic case, i.e.,~$J>0$, as~stated above. Now energy difference between lowest two energy levels is $8J-2B$ and highest two energy levels is $2B$. Thus, for~a fixed $J$, with~increasing $B$, spacing between lowest two energy levels decreases, where       spacing between highest two energy level increases. Thus, as~a whole, we cannot say that the adiabatic stroke is compression or expansion for the working medium and consequently the line of reasoning   in  \cite{talkner1} does not hold well here. We simply say that the work strokes are the first and second adiabatic work~strokes.

\begin{figure}[h]
\centering
%\begin{center}
%\begin{minipage}{0.3\linewidth}
\includegraphics[height=3.5cm,width=4.6cm]{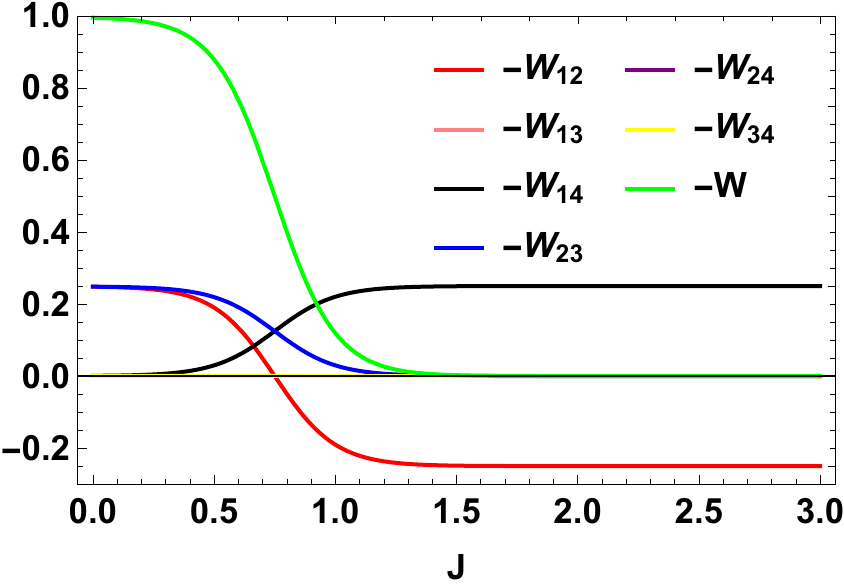}
\caption{(Color online) $-W_{mn}$ and $-W$ vs. $J$ plot for $B_2>B_1$, with~$B_2=4$, $B_1=3$.}
\label{spin1contri2}
%\end{minipage}
%\end{center}
\end{figure}

\begin{figure}[h]
\centering
%\begin{center}
%\begin{minipage}{0.3\linewidth}
\includegraphics[height=3.5cm,width=4.6cm]{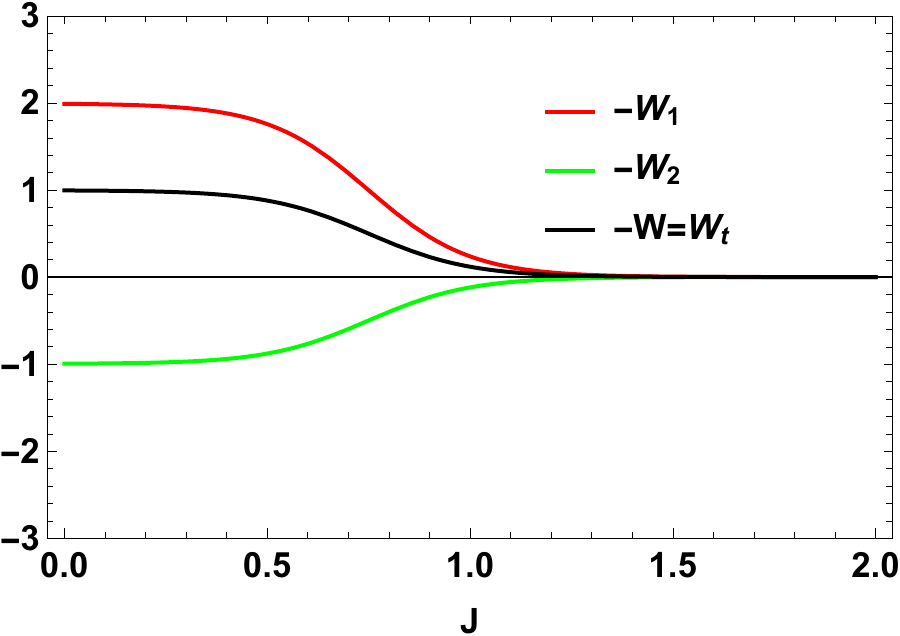}
\caption{(Color online) $-W_1$, $-W_2$ and $W_t$ vs. $J$ plot for $B_2>B_1$, with~$B_2=4$, $B_1=3$. }
\label{spin1contri1}
%\end{center}
\end{figure}

\begin{figure}[h]
\centering
%\begin{center}
%\end{minipage}
%\hfill
%\begin{minipage}{0.3\linewidth}
\includegraphics[height=3.5cm,width=4.6cm]{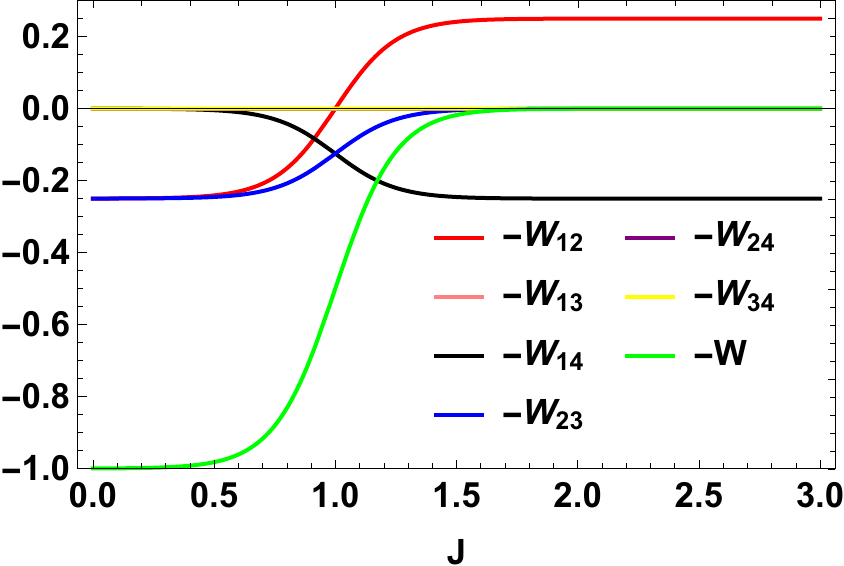}
\caption{(Color online) $-W_{mn}$ and $-W$ vs. $J$ plot for $B_2<B_1$, with~$B_2=3$, $B_1=4$. }
%\includegraphics[height=5cm,width=7.5cm]{wcon1.eps}
%\caption{ }
\label{spin1contri}
%\end{minipage}
%\end{center}
\end{figure}

Thus, whether the extracted work is positive or negative would depend upon the structure of the energy levels of the Hamiltonian. Let us  illustrate this by considering the working medium to be a two spin-1/2 system. 
%\begin{figure}[h]
%\begin{center}
%\hfill
%\includegraphics[height=5cm,width=7.5cm]{wcon1.eps}
%\caption{(Color online) Work contributions ($-W_{mn}$) and total extracted work ($-W$) vs $J$ plot for $B_2>B_1$, with $B_2=4$, $B_1=3$. }
%\label{spin1contri2}
%\end{center}
%\end{figure}
In Equation~(\ref{negwork1}), some terms are negative and some terms are positive. Sign of the total work depends upon the positive and negative contributions of these terms. We denote each term of the expression as $W_{mn}=\frac{1}{2}(\Delta^f_{m,n}-\Delta^i_{m,n})T_{m,n}[P_m(B_1)-P_n(B_1)]$ and note that $W_{mn}=W_{nm}$. We~  plot  different terms with the coupling constant $J$ in Figure~\ref{spin1contri2} for $B_2>B_1$. We notice that  $-W_{12}$ ($-W_{21}$) gives negative contributions, $-W_{14}$ ($-W_{41}$) and $-W_{23}$ ($-W_{32}$) give positive contributions and the other terms are zero. The~balances between these terms   decide the sign of the work output. In~this case, it is positive.
We also plot the contributions for first and second adiabatic work strokes in Figure~\ref{spin1contri1} for $B_2>B_1$, and~notice that  the~contribution of the first work stroke is positive, whereas the contribution of the second work stroke is negative but their sum is positive and hence we extract  positive work.
It is important to note that the  arguments behind $Q_M\geq 0$ and $Q_T\leq 0$ are independent of the choice of initial and final magnetic field $B_1$ and $B_2$, respectively.  Indeed, for both $B_2>B_1$ and $B_2<B_1$, $Q_M$ is always positive and $Q_T$ is always negative. It is the extracted work $W_t$, whose sign depends upon $B_1$ and $B_2$. From~Equation~(\ref{wglosphalf}), it is evident that, when $B_2>B_1$, we extract positive work, i.e.,~$W_t\geq 0$ and, when $B_2<B_1$, extracted work is negative for all $J$. This can be seen  in      Figure~\ref{spin1contri}. This plot shows that when the values of $B_1$ and $B_2$ are interchanged, i.e.,~$B_1=4$ and $B_2=3$, the~nature of the plots are just opposite to each other. 
From the definition of efficiency as { the ratio of total work extracted ($W_t$) over $Q_M$}, $\eta=W_{t}/Q_M$, it is evident that the efficiency can be negative only if the extracted work $W_t$ is negative, { which means that we cannot extract work from the engine but have to do work to run it}. We have seen that, for $B_2>B_1$, the~extracted work is positive. For~the rest of the analysis, we   stick to this scenario, i.e.,~$B_2>B_1$.
From the expressions derived above, we have different efficiencies depending upon the measurement choices. When $\vec{\sigma}.\hat{a}$ is $\sigma_z$ and $\vec{\sigma}.\hat{b}$ is $\sigma_x$, we have,
\begin{equation}
\eta=\frac{(B_1-B_2)(1-e^{4B_1})}{B_2(-1+e^{4B_1})-2[1+e^{2B_1}(1+e^{2B_1}-3e^{8J})]J}.
\label{effi1}
\end{equation}
%\begin{figure}[H]
%\begin{center}
%\includegraphics[height=4cm,width=6cm]{zx.eps}
%\caption{ Efficiency vs $J$ plot for $\sigma_z$ and $\sigma_x$}
%\label{xy}
%\end{center}
%\end{figure}

Thus, the efficiency depends upon the coupling constant $J$, where       $J=0$ corresponds to the uncoupled scenario with efficiency $(1-\frac{B_1}{B_2})$. As~$J$ increases, for~a certain range of $J$, efficiency is also increased over the uncoupled value. Expression for the efficiency remains same as in Equation~(\ref{effi1}) for $\vec{\sigma}.\hat{a}=\sigma_y$ and $\vec{\sigma}.\hat{b}=\sigma_z$,
For, $\vec{\sigma}.\hat{a}$=$\sigma_x$ and $\vec{\sigma}.\hat{b}$=$\sigma_y$,
\begin{equation}
\eta=\frac{(B_1-B_2)(1-e^{4B_1})}{B_2(-1+e^{4B_1})-[1+e^{2B_1}(1+e^{2B_1}-3e^{8J})]J}.
\end{equation}
\vspace{-12pt}
\unskip
\unskip
%\begin{figure}[h]
%\begin{center}
%\includegraphics[height=5cm,width=7.5cm]{wcon1wef.pdf}
%\hfill
%\includegraphics[height=5cm,width=7.5cm]{wcon1.eps}
%\caption{(Color online) Work contributions ($-W_{mn}$) and total extracted work ($-W$) vs $J$ plot for $B_2<B_1$, with $B_2=3$, $B_1=4$. }
%\label{spin1contri}
%\end{center}
%\end{figure}

Now, let us examine those cases where same observables are being measured on both sides, e.g.~$\vec{\sigma}.\hat{a}=\vec{\sigma}.\hat{b}=\sigma_z$, $\sigma_x$ or $\sigma_y$. When both   observables are $\sigma_x$, we have,
\begin{equation}
\eta=\frac{(B_1-B_2)(-1+e^{4B_1})}{B_2(1-e^{4B_1})+(1+e^{4B_1}-2e^{2B_1+8J})J}.
\end{equation}

\begin{figure}[h]
\centering
%\begin{center}
%\begin{minipage}{0.45\linewidth}
\includegraphics[scale=.72]{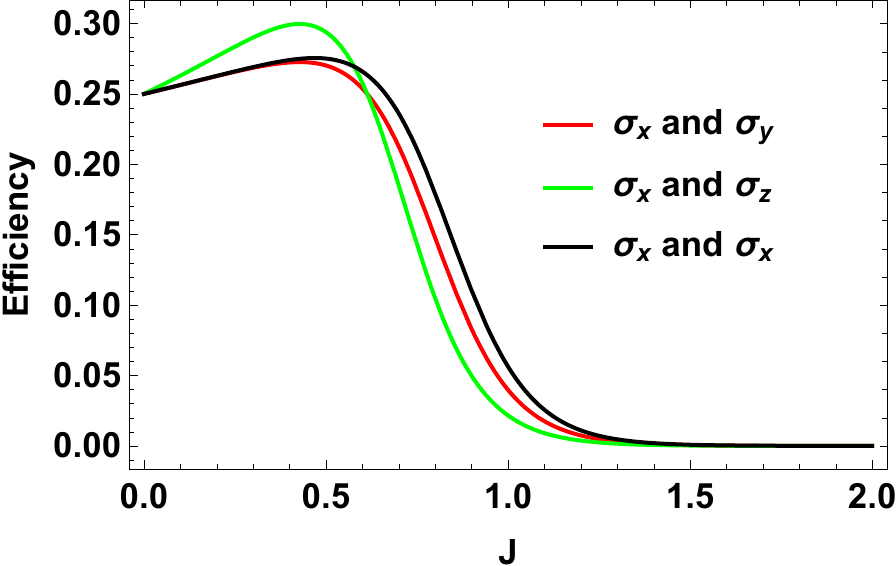}
\caption{(Color online) Efficiency vs. $J$ plot for different measurement choices for two spin-$1/2$~particles.}
\label{spinhalf1}
%\end{minipage}
\end{figure}

\begin{figure}[h]
\centering
%\hfill
%\begin{minipage}{0.45\linewidth}
\includegraphics[height=5cm,width=6cm]{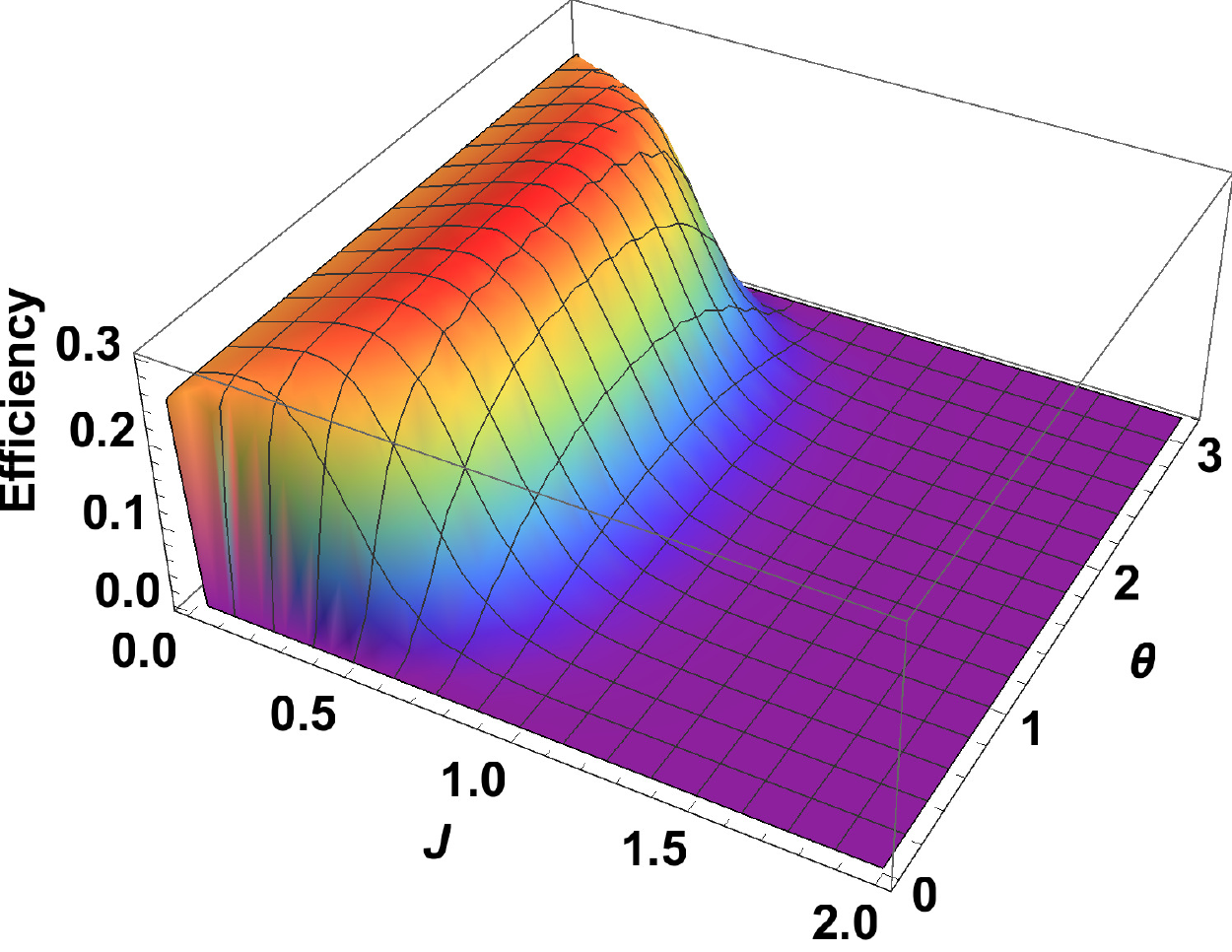}
\caption{(Color online) Efficiency vs. $J$ plot for $\sigma_z$ on one side and arbitrary observable (depends upon $\theta$) on other~side.}
\label{xy}
%\end{minipage}
%\end{center}
\end{figure}

Exactly the same expression is obtained when both   observables are $\sigma_y$.~When both   observables are $\sigma_z$, the~work contributions from two adiabatic branches are equal and opposite to each other. Consequently, the~total work done is zero and hence the efficiency is zero.
After calculating the efficiency for different measurement choices, we plot them in Figure~\ref{spinhalf1}  together  with  $B_1=3$ and $B_2=4$. Next, we   plot  the efficiency for a fixed observable $\sigma_z$ on one side and varying the parameters for the observable on the other side. We can write  $\vec{\sigma}.\hat{m}=\sin\theta\cos\phi\sigma_x+\sin\theta\sin\phi\sigma_y+\cos\theta\sigma_z$. In~this case, we calculate the efficiency and it turns out to be independent of the parameter $\phi$ but depends on $\theta$.               
From the 3D plot in Figure~\ref{xy}, it is clear that, when $\theta$ is $\pi/2$, the~efficiency is optimum and it is exactly equal to the case where $\sigma_x$ is measured for one spin and $\sigma_z$ is measured for the~other. Our~results show that using non-zero coupling $J$, we can actually get an advantage over the no-coupling scenario.
We~  argue  that, for~$B_2>B_1$, the extracted work is positive and hence we get a positive efficiency. Now, we   show that, for any $B_1$ and $B_2$ ($B_1, B_2>0$) with $B_2>B_1$, the~efficiency of the coupled engine is greater than that of an uncoupled one, for~a certain range of $J$. To~show this, we~rewrite Equation~(\ref{effi1})~as,
\begin{eqnarray}
&\eta=\Big(1-\frac{B_1}{B_2}\Big)\Big[\frac{(-1+e^{4B_1})}{(-1+e^{4B_1})-2B_2[1+e^{2B_1}(1+e^{2B_1}-3e^{8J})]J}\Big].
\label{effi2}
\end{eqnarray}
\vspace{-12pt}

As already mentioned, $\Big(1-\frac{B_1}{B_2}\Big)$ is the efficiency of an uncoupled ($J=0$) engine.
From the above expression, it turns out that the efficiency will be greater than that of the uncoupled one if $2B_2[1+e^{2B_1}(1+e^{2B_1}-3e^{8J})]J>0$. This implies that,
\begin{equation}
\label{valuecal}
e^{2B_1}(3e^{8J}-e^{2B_1}-1)<1.
\end{equation}

From this inequality, it is evident that for any $B_1>0$ ({ $B_2$ can have any value greater than $B_1$}), we always get a positive value for $J$, below~which we get an advantage for the efficiency over the uncoupled one. To~give an example, let   us consider a small value of $B_1$, e.g.,~$B_1=0.1$. When $B_1=0.1$, we have to find the value of $J$ for which the above inequality holds. { Solving the Equation~(\ref{valuecal}) for $B_1=0.1$}, one can show that when $J<0.00166$ approximately, the~efficiency will be greater than that for the uncoupled one. For~$B_1=0.1$, { and $B_2=4$}, the~uncoupled efficiency is 0.975. Let us  take $J=0.0014$. For~that, we have $\eta=0.975011$ { for the same values of $B_1$ and $B_2$, i.e.,~$B_1=0.1$ and $B_2=4$.} Thus, the~coupled engine is still more efficient but it is in such a small region of that nature of $J$ that cannot be seen from the plot unless the plot has a very fine scaling. { With the increase in the value of $B_1$, the~cutoff value of $J$, above~which the coupled engine is more efficient increases and   the lower is the value of the ratio of $B_1$ over $B_2$,     the greater is the  efficiency of the engine. }
For the uncoupled efficiency       closer and closer  to one, the~region in which there is an advantage of the coupled engine will be     narrower and narrower. However, in~principle, we always have the coupled engine as the efficient one, compared to the uncoupled one.
In addition,  in~Figure~\ref{spinhalf1}, it is evident that different measurement choices give different efficiencies for the heat engine. Here, choosing $\sigma_x$ and $\sigma_z$ or $\sigma_y$ and $\sigma_z$ as measurement operators, we get the maximum efficiency. Thus, judiciously choosing measurement operators is important for optimum performance of the heat engine.
%\begin{figure}[H]
%\begin{center}
%\includegraphics[height=5cm,width=8.5cm]{plot1arb.pdf}
%\caption{(Color online) Efficiency vs $J$ plot for $\sigma_z$ on one side and arbitrary observable (depends upon $\theta$) on other side.}
%\label{xy}
%\end{center}
%\end{figure}
\section{Higher-Dimensional~Case}
\label{sec6}
In this section, we are interested in the higher-dimensional Heisenberg model as considered in~\cite{ferdi}, where one spin half particle is coupled to a spin $s$ particle. It would  be interesting to observe the effect of higher spin as an additional parameter along with the coupling constant.~We have the system~Hamiltonian,\vspace{6pt}
\begin{equation}
H=8J\vec{S_A}.\vec{S_B}+2B(S_A^z+S_B^z),
\label{hamil}
\end{equation}
where       $\vec{S_A}=(S_A^x,S_A^y,S_A^z)$ and $\vec{S_B}=(S_B^x,S_B^y,S_B^z)$ are two spin operators, $J$ is the system-bath coupling constant and $B$ is the external magnetic field. Two spin-$1/2$ cases have  already been discussed. Now, the detailed calculation   is carried out for $S_{A/B}=1$ and $S_{A/B}=3/2$. One can obviously go on to calculate the cases for $2$ and $5/2$ and so on, but~the essential points can be observed by studying the following cases: $S_{A/B}=1$ and $S_{A/B}=3/2$. We first start with a spin-$1/2$ and a spin-$1$ operator, which is an asymmetric case in the sense that the spins on two sides are different. We call it symmetric when two spins at two sides are the same. We   deal with  the different cases one by one. We start with the asymmetric~cases.
\subsection{Asymmetric~Case}%: $s_A=1/2$ and $S_B=1$}
\begin{figure}[h]
\centering
%\begin{center}
%\begin{minipage}{0.3\linewidth}
\includegraphics[height=3.5cm,width=4.6cm]{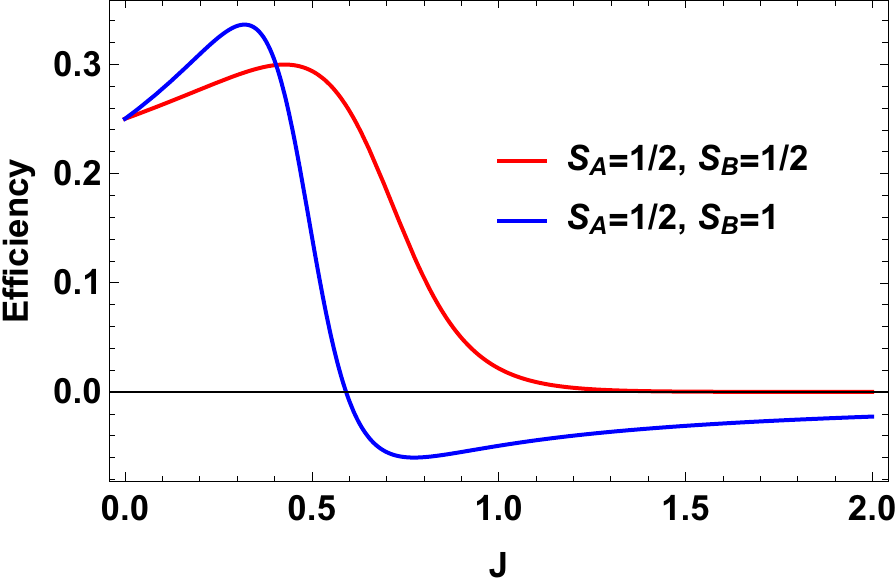}
\caption{(Color online) Efficiency vs. $J$ plot for $B_2=4$ and $B_1=3$.}
\label{spin1}
\end{figure}

\begin{figure}[h]
\centering
%\end{minipage}
%\hfill
%\begin{minipage}{0.3\linewidth}
\includegraphics[height=3.5cm,width=4.6cm]{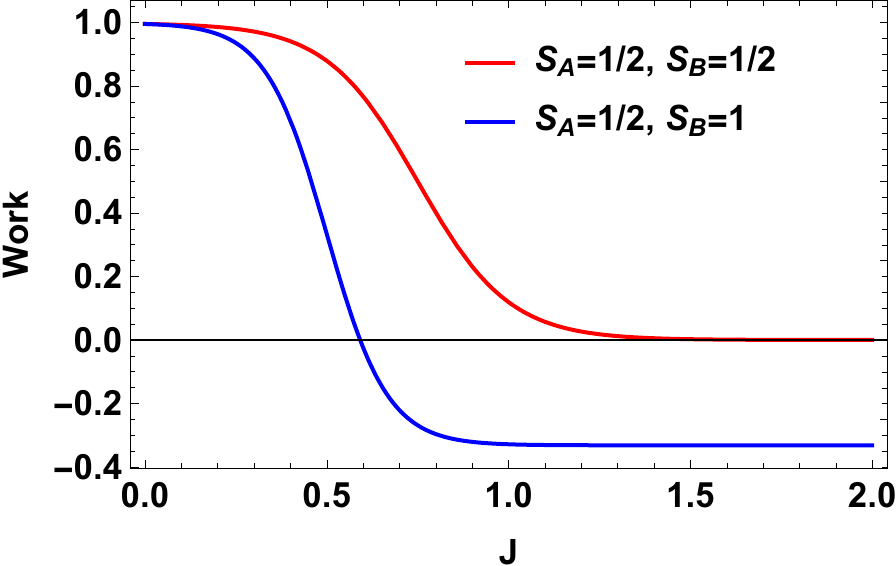}
\caption{(Color online) Total work vs. $J$ plot for  $B_2=4$ and $B_1=3$.}
\label{work}
%\end{minipage}
\end{figure}

\begin{figure}[h]
\centering
%\hfill
%\begin{minipage}{0.3\linewidth}
\includegraphics[height=3.5cm,width=4.6cm]{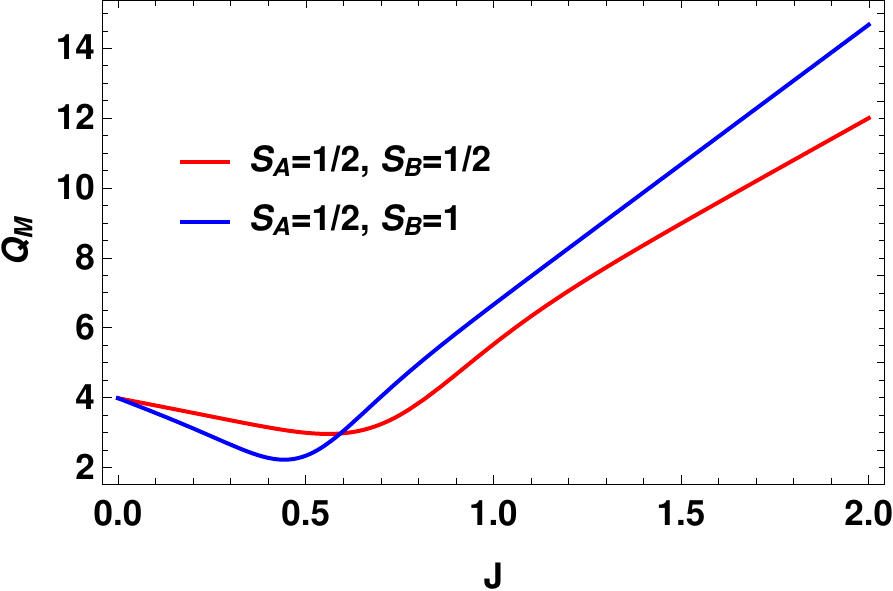}
\caption{(Color online) Average energy change $Q_M$  vs. $J$ plot for  $B_2=4$ and $B_1=3$.}
\label{heat}
%\end{minipage}
%\end{center}
\end{figure}

First, we take $S_A=1/2$ and $S_B=1$.~Spin operators for spin-$1/2$ particle are $\frac{1}{2}\vec{\sigma}$, where~$\vec{\sigma}=(\sigma_x,\sigma_y,\sigma_z)$ are the Pauli matrices. These spin operators for spin-$1/2$ particle form the fundamental irreducible representation of $SU(2)$. Spin operators for the spin $1$ particle are the three-dimensional irreducible representation of $SU(2)$. With~standard representation in canonical bases~\cite{georgi}, %These are listed as following, 
%\[
%S^x=\frac{1}{\sqrt{2}}
 % \begin{pmatrix}
  %  0 & 1 & 0 \\
   % 1 & 0 & 1 \\
    %0 & 1 & 0
  %\end{pmatrix},
%S^y=\frac{1}{\sqrt{2}}
 % \begin{pmatrix}
  %  0 & -i & 0 \\
   % i & 0 & -i \\
   % 0 & i & 0
  %\end{pmatrix},
%\]  
%\[
%S^z=
 % \begin{pmatrix}
  %  1 & 0 & 0 \\
   % 0 & 0 & 0 \\
    %0 & 0 & -1
  %\end{pmatrix}.\]
%With these spin matrices, 
the~eigenvalues and the eigenstates of the Hamiltonian given in Equation~(\ref{hamil}) are listed in Table~\ref{tabhalfone}. 
%\unskip
%where       $\ket{0_A}\doteq
%\begin{pmatrix} 1\\0\end{pmatrix}$, $\ket{1_A}\doteq
%\begin{pmatrix} 0\\1\end{pmatrix}$, $\ket{0_B}\doteq
%\begin{pmatrix} 1\\0\\0\end{pmatrix}$, $\ket{1_B}\doteq
%\begin{pmatrix} 0\\1\\0\end{pmatrix}$ and $\ket{2_B}\doteq
%\begin{pmatrix} 0\\0\\1\end{pmatrix}$, are a set basis vectors for the two and three-dimensional Hilbert spaces, respectively and they are the eigenstates of $S^z$ operator both for $A$ and $B$ side. 
As in the previous cases, for~the measurement step of the engine cycle, we have a number of choices for the measurement operators and we explore  those options in the previous sections for the case of two spin-$1/2$ particles. Now, for~brevity, we     consider  one particular measurement setup and observe the effect of higher spin, such that this spin can also be a controlling parameter for the efficiency.
We choose the measurement operators as,
\begin{eqnarray}
\nonumber
&M_1=\ket{0^x_A}\bra{0^x_A}\otimes \ket{0_B}\bra{0_B}, M_2=\ket{0^x_A}\bra{0^x_A}\otimes \ket{1_B}\bra{1_B} \\
\nonumber
&M_3=\ket{0^x_A}\bra{0^x_A}\otimes \ket{2_B}\bra{2_B}, M_4=\ket{1^x_A}\bra{1^x_A}\otimes \ket{0_B}\bra{0_B} \\
\nonumber
&M_5=\ket{1^x_A}\bra{1^x_A}\otimes \ket{1_B}\bra{1_B}, M_6=\ket{1^x_A}\bra{1^x_A}\otimes \ket{2_B}\bra{2_B},
\end{eqnarray}
where       $\ket{0^x_A}=\sqrt{1/2}[\ket{0}+\ket{1}]$ and $\ket{1^x_A}=\sqrt{1/2}[\ket{0}-\ket{1}]$ are the eigenstates of the operator $S^x$ for the spin-half particle $A$. In~other words, we are doing measurement of the operator $S^x$ on side $A$ for the spin-$1/2$ and measurement of the operator $S^z$ on the side $B$ for spin-$1$. In~the plots (see~Figures~\ref{spin1}--\ref{heat}), we compared different scenarios for the same measurement settings on the two sides.~By~the same measurement settings, we mean that on the spin half side the measurement operators will be the projectors constructed from the eigenstates of the operator $S^x$ and, on the higher spin side, it will be the projectors of the eigenstates of $S^z$.
For the above measurement operators, we calculate the quantities as work and heat (during measurement process and thermalization step) and evaluate the efficiency of the heat engine. The~expression for the total work extracted for this case is,
\begin{eqnarray}
\nonumber
&&W_t=-W=\\
&&\frac{(B_2-B_1)(-1+e^{2B_1})(3 + e^{2 B1}(4 + 3 e^{2 B1} - e^{12 J}))}{3(1 + e^{2 B1}) (1 + e^{4 B1} + e^{2(B_1 + 6 J)})}
\end{eqnarray}
\vspace{-12pt}
\begin{table}[h]
\centering
\caption{$S_A=1/2$,$S_B=1$.}\label{tabhalfone}
\begin{tabular}{cccc}
%\hline
%Acetylene: hydrogen concentration & \multicolumn{3}{|r|}{Average roughness values $S_a$ in nm} \\[0.5 ex]
%\hline
%{} &{|r|}{Value of $p$ to start violation} \\[0.3 ex]
\toprule
%{} & {}\\
%{} &\multicolumn{2}{|r|}{Value of $p$ to start violation} \\[0.3 ex]
%\hline
%\cline{1-2}
{\textbf{Eigenvalues}}  &   \textbf{Eigenstates} \\
%\midrule
\hline
$-B-8J=E_1$ &$-\sqrt{\frac{2}{3}}\ket{0_A2_B}+\sqrt{\frac{1}{3}}\ket{1_A1_B}=\ket{\psi_1}$\\

$B-8J=E_2$ & $-\sqrt{\frac{1}{3}}\ket{0_A1_B}+\sqrt{\frac{2}{3}}\ket{1_A0_B}=\ket{\psi_2}$\\

$-3B+4J=E_3$ & $\ket{1_A2_B}=\ket{\psi_3}$\\

$-B+4J=E_4$ & $\sqrt{\frac{1}{3}}\ket{0_A2_B}+\sqrt{\frac{2}{3}}\ket{1_A1_B}=\ket{\psi_4}$\\

$B+4J=E_5$ & $\sqrt{\frac{2}{3}}\ket{0_A1_B}+\sqrt{\frac{1}{3}}\ket{1_A0_B}=\ket{\psi_5}$\\

$3B+4J=E_6$ & $\ket{0_A0_B}=\ket{\psi_6}$\\
%\bottomrule
\hline
\end{tabular}
\end{table}

\begin{figure}[h]
\centering
%\begin{center}
%\begin{minipage}{0.3\linewidth}
\includegraphics[height=3.5cm,width=4.6cm]{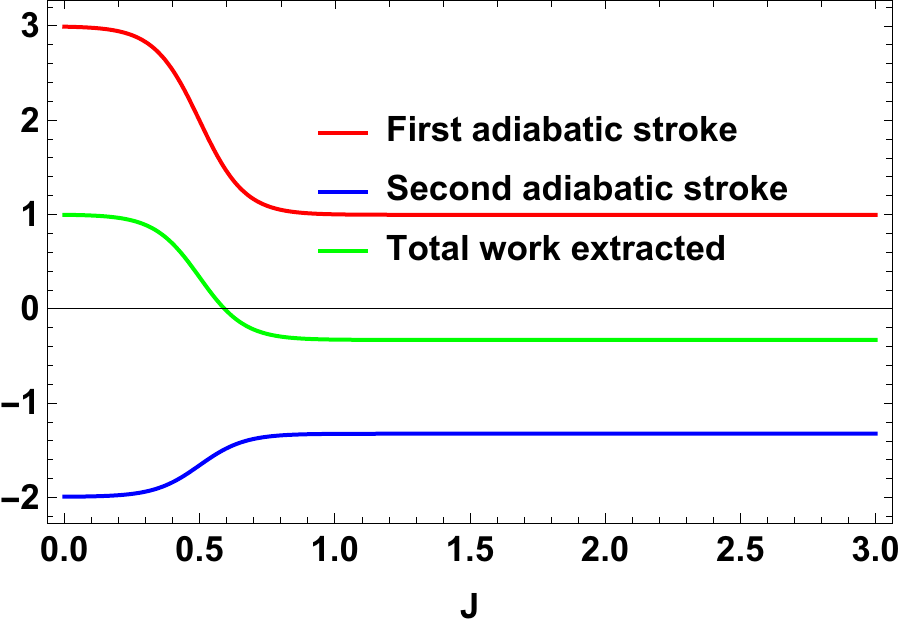}
\caption{(Color online) $-W_1$, $-W_2$, $W_t$ vs. $J$ plot for $B_2>B_1$, with~$B_2=4, B_1=3$ ; $S_A=1/2$, $S_B=1$.}
\label{negv1}
%\end{minipage}
\end{figure}

\begin{figure}[h]
\centering
%\vspace{0.00mm}
%\hfill
%\begin{minipage}{0.3\linewidth}
\includegraphics[height=3.5cm,width=4.6cm]{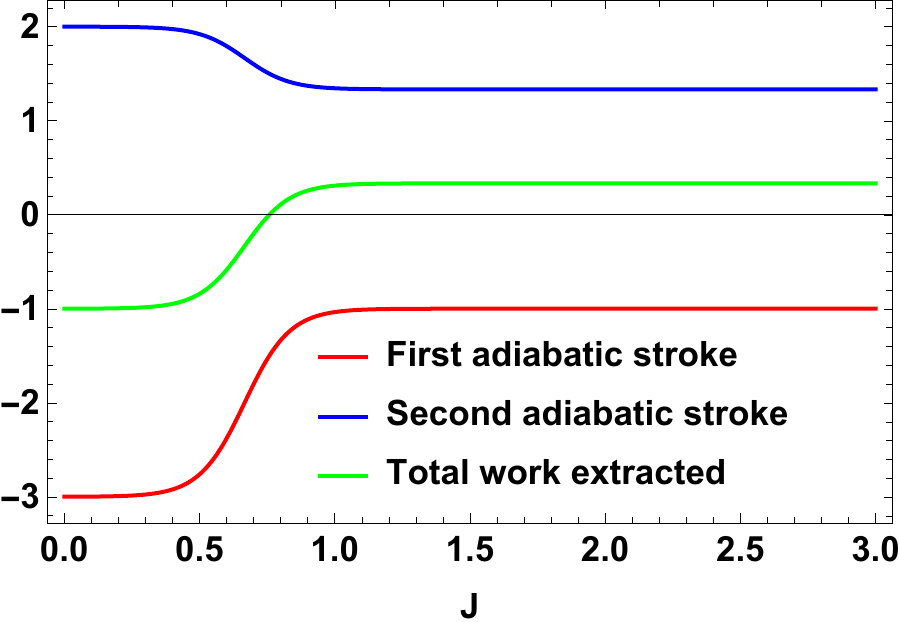}
\caption{(Color online) $-W_1$, $-W_2$, $W_t$ vs. $J$ plot for $B_2<B_1$, with~$B_2=3, B_1=4$ ; $S_A=1/2$, $S_B=1$.}
\label{negv11}
\end{figure}

\begin{figure}[h]
\centering
%\end{minipage}
%\hfill
%\begin{minipage}{0.3\linewidth}
\includegraphics[height=3.5cm,width=4.6cm]{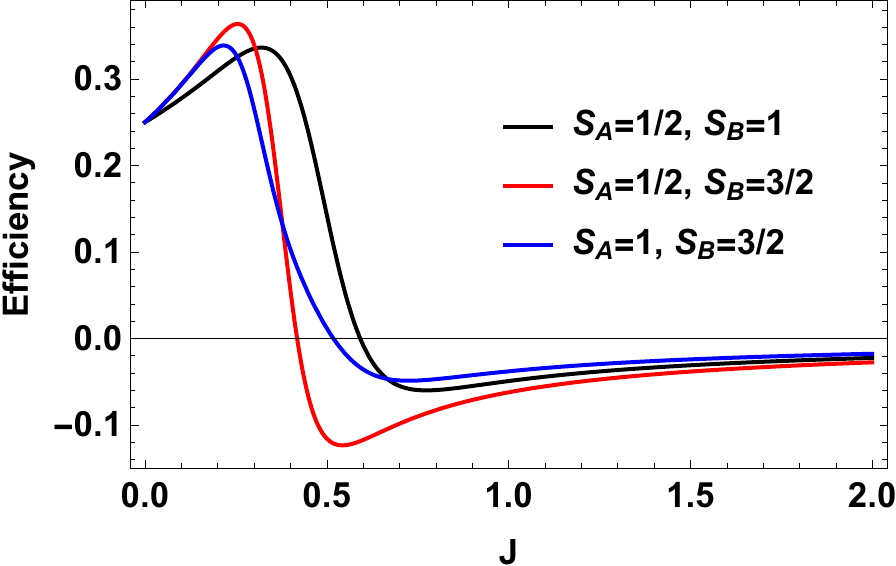}
\caption{(Color online) Efficiency vs. $J$ plot for for $B_2>B_1$, with~$B_2=4, B_1=3$ for asymmetric cases.}%($S_A=1/2$, $S_B=1$), ($S_A=1/2$, $S_B=3/2$) and ($S_A=1$, $S_B=3/2$) ; $B_2=4$ and $B_1=3$.}
\label{plot1sp13}
%\end{minipage}
%\end{center}
\end{figure}

\begin{figure}[h]
\centering
%\begin{center}
%\begin{minipage}{0.32\linewidth}
\includegraphics[height=3.5cm,width=4.3cm]{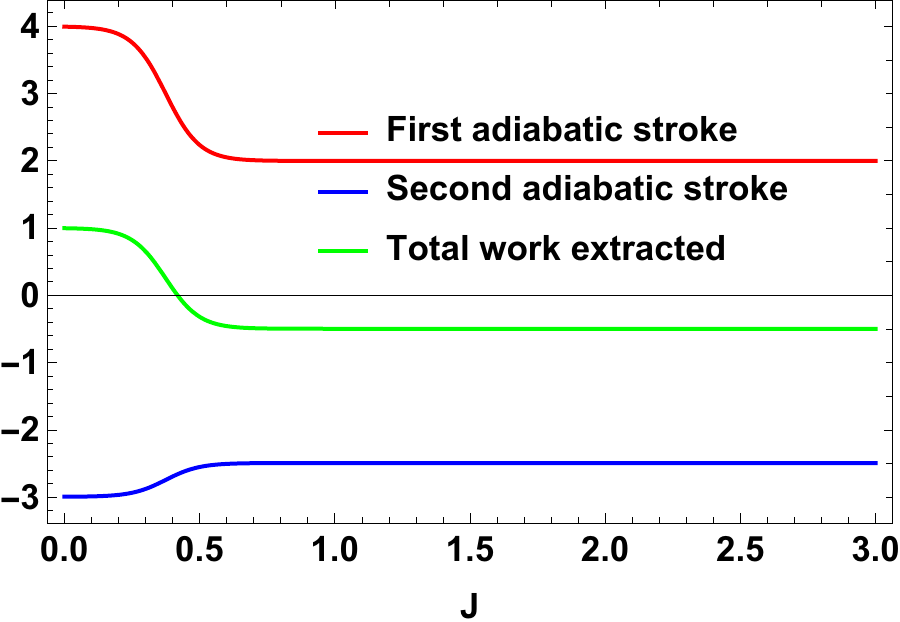}
\caption{(Color online) $-W_1$, $-W_2$, $W_t$ vs. $J$ plot for $B_2>B_1$, with~$B_2=4, B_1=3$. Here, $S_A=1/2$, $S_B=3/2$. }%This figure has now been referred in section 5.1. Earlier the lables got identical with other figures.
\label{highspinwork1}
\end{figure}

\begin{figure}[h]
\centering
%\end{minipage}
%\hfill
%\begin{minipage}{0.32\linewidth}
\includegraphics[height=3.5cm,width=4.4cm]{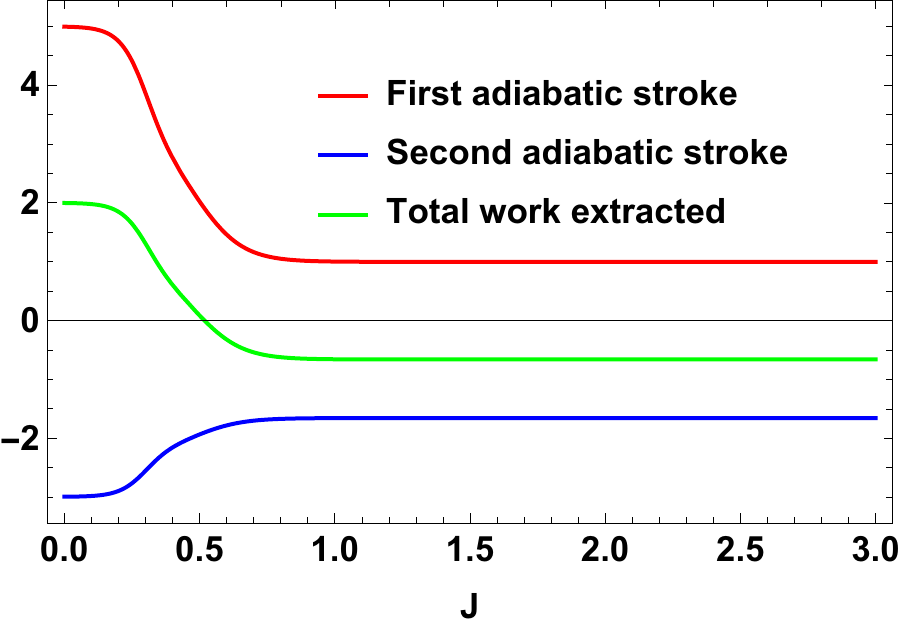}
\caption{(Color online) $-W_1$, $-W_2$, $W_t$ vs. $J$ plot for $B_2>B_1$, with~$B_2=4, B_1=3$. Here, $S_A=1$, $S_B=3/2$.}%This figure has now been referred in section 5.1. Earlier the lables got identical with other figures.
\label{highspinwork2}
\end{figure}

From the expression above, one can see that only $B_2>B_1$ will not guarantee the positivity of $W_t$ for all $J$, unlike the previous case of two spin-$1/2$ particles. Specifically, the~work extracted is negative when $4 + 3 e^{2 B1}-e^{12 J}<0$. This is not the case for two spin-$1/2$ particles, where expression of the extracted work is such that $B_2>B_1$ always gives positive extracted work and $B_1>B_2$ gives negative extracted work. However,~in the present case, for~$B_2>B_1$, extracted work can be negative and, for $B_1>B_2$, extracted work can be positive when  $4 + 3 e^{2 B1}-e^{12 J}<0$. As~we     show below,   this is special for the asymmetric cases only. In~Figure~\ref{spin1}, we plot the efficiency and compare it with the spin half scenario for $B_2>B_1$, with~same values of magnetic fields previously considered, i.e, $B_1=3$ and $B_2=4$. 
We note that (Figure \ref{spin1}) the efficiency for the spin-$1$ scenario can be higher than that of spin half scenario for some range of non-zero values of $J$. For~the uncoupled case, both   engines give same efficiencies, which is $1-\frac{B_1}{B_2}$. 
Another point to note is that the efficiency can go to negative for the spin $1$ scenario. As~discussed  above, the~negative efficiency comes entirely from the negative { extracted} work, because~$Q_M$ is always positive. The~total { extracted} work $W_t$ can now be negative starting from a certain value of $J$, as shown in Figure~\ref{work}, for $B_2>B_1$, with~$B_2=4$ and $B_1=3$, whereas $Q_M$ is always positive, as  shown in Figure~\ref{heat}.
Now, one can further analyze the work strokes and see which work stroke is contributing more for negative work.  From the {Figure}~\ref{negv1}, one can note that, as before, the~first adiabatic work stroke gives the positive work output, whereas, for second adiabatic work stroke, we get negative work. 
%\begin{figure}[h]
%\begin{center}
%\includegraphics[height=5cm,width=7.5cm]{work-n.pdf}
%\caption{(Color online) Global work vs $J$ plot for ($S_A=1/2$, $S_B=1/2$) and ($S_A=1/2$, $S_B=1$) ; $B_2=4$ and $B_1=3$.}
%\label{work}
%\end{center}
%\end{figure}
%\begin{figure}[H]
%\begin{center}
%\includegraphics[height=5cm,width=7.5cm]{heat-n.pdf}
%\caption{(Color online) Average energy change $Q_M$ during the measurement step vs $J$ plot for ($S_A=1/2$, $S_B=1/2$) and ($S_A=1/2$, $S_B=1$) ; $B_2=4$ and $B_1=3$.}
%\label{heat}
%\end{center}
%\end{figure}
If we take $B_1>B_2$, the~situation is reversed. 
We have shown the case of $B_1>B_2$, with~$B_1=3$ and $B_2=4$, in~Figure~\ref{negv11}. The~plot is exactly opposite to the previous one with $B_2=4$ and $B_1=3$ (see {Figure}~\ref{negv1}). 
Thus, in~the range of $J$ where the efficiency is negative, the~situation appears as follows: the average energy $Q_M$ is entering to the working medium, $W_t$ work is being done and thereby heat $Q_T$ goes into the heat bath of temperature $T$. In~some sense, one can associate a refrigerator action for this negative work scenario. In~a conventional quantum Otto refrigerator, the~system is first prepared in the thermal state with temperature corresponding to the cold bath. Then, in the first and third adiabatic strokes, a total work $W$ is added to the working medium. In~the second and fourth steps, $Q_2$ heat is taken from the cold bath and $Q_1$ heat is added to the hot heat bath and eventually cooling the cold bath more. The~co-efficient of performance (COP) for the refrigerator is given as $\eta_{COP}=Q_2/W$. In~our case, $Q_M$ is always positive and $Q_T$ is always negative. This indicates that we can consider a cold bath of effective temperature $T_2$, such that, after~the second stroke, we have,%Figure 10 should be cited after      Figure 9, please renumber      Figures so they appear in sequential numerical order.--- I have added a small part above (highlighted in blue color on page 12), where Figure 9 is cited before the Figure 10.
\begin{equation}
Q_M=Tr[((\rho^{eq}(T,B_2)-\rho^{eq}(T_2,B_2))H(B_2)],
\end{equation}
where       $\rho^{eq}(T',B_2)$ is the thermal state at temperature $T'$ with the Hamiltonian $H(B_2)$, given in Equation~(\ref{hamil}), where $T'=(T,T_2)$. Solving the above equation we can associate an effective temperature $T_2$ with $Q_M$. Thus, now one can read the cycle as  transferring heat from the cold bath of temperature $T_2$ to hot bath of temperature $T$ and, for this, $W_t$ work has to be done. This means that the COP of this refrigerator action is $Q_M/(-W_{t})$.
%That means that work can not be extracted but we have to do work on the system for the cycle to run. It's more like a refrigerator without the cold bath from which we draw heat for a refrigerator. Instead, that step is replaced by a measurement step, which changes the average energy of the system. 
Thus,   to get a positive work output, we have to judiciously choose the value of the coupling constant $J$, such that efficiency is not negative. Then, we   get an advantage for higher efficiency over the uncoupled one.
Let us  now consider the next asymmetric scenarios and see whether a similar trend, i.e, increase in efficiency and occurrence of negative efficiency is present or not. 
%\subsection{Asymmetric scenario} %$s_A=1/2$, $S_B=3/2$.}
We start with the case where the spins on two sides are $S_A=1/2$ and  $S_B=3/2$. %and $s_A=1$, $S_B=3/2$.
With~standard representation of spin-$3/2$ in canonical bases~\cite{georgi},
%For spin $3/2$, we have the following spin operators,
%\[
%S^x=\frac{1}{2}
 % \begin{pmatrix}
  %  0 & \sqrt{3} & 0 & 0 \\
   % \sqrt{3} & 0 & 2 & 0\\
    %0 & 2 & 0 & \sqrt{3}\\
    %0 & 0 & \sqrt{3} & 0
  %\end{pmatrix},
%S^z=\frac{1}{2}
 % \begin{pmatrix}
  %  3 & 0 & 0 & 0 \\
   % 0 & 1 & 0 & 0\\
   % 0 & 0 & -1 & 0\\
    %0 & 0 & 0 & -3
  %\end{pmatrix},
%\]
%and
%\[
%S^y=\frac{1}{2}
 % \begin{pmatrix}
  %  0 & -i\sqrt{3} & 0 & 0 \\
   % i\sqrt{3} & 0 & -2i & 0\\
    %0 & 2i & 0 & -i\sqrt{3}\\
    %0 & 0 & i\sqrt{3} & 0
  %\end{pmatrix}.
%\]
the eigenvalues and eigenstates of the Hamiltonian in Equation~(\ref{hamil}) are given in   Appendix \ref{app1} in Table~\ref{tabhalf3}.
Now, we choose the same kind of   measurement operators  as in  the previous case, i.e.,~on the spin-$1/2$ side we measure $S_x$ and on the spin-$3/2$ side, $S_z$:
\begin{eqnarray}
\nonumber
&M_1=\ket{0^x_A}\bra{0^x_A}\otimes \ket{0_B}\bra{0_B}, M_2=\ket{0^x_A}\bra{0^x_A}\otimes \ket{1_B}\bra{1_B} \\
\nonumber
&M_3=\ket{0^x_A}\bra{0^x_A}\otimes \ket{2_B}\bra{2_B}, M_4=\ket{0^x_A}\bra{0^x_A}\otimes \ket{3_B}\bra{3_B} \\
\nonumber
&M_5=\ket{1^x_A}\bra{1^x_A}\otimes \ket{0_B}\bra{0_B}, M_6=\ket{1^x_A}\bra{1^x_A}\otimes \ket{1_B}\bra{1_B} \\
\nonumber
&M_7=\ket{1^x_A}\bra{1^x_A}\otimes \ket{2_B}\bra{2_B}, M_8=\ket{1^x_A}\bra{1^x_A}\otimes \ket{3_B}\bra{3_B}, %\\
%\nonumber
\end{eqnarray}%\unskip
with $\ket{0^x_A}=\sqrt{1/2}(\ket{0_A}+\ket{1_A})$, $\ket{1^x_A}=\sqrt{1/2}(\ket{0_A}-\ket{1_A})$, and~$\ket{0_B}$, $\ket{1_B}$, $\ket{2_B}$, and $\ket{3_B}$ are the eigenstates of $S_B^z$ corresponding to the eigenvalues 3, 1, $-$1, and $-$3, respectively.
We also consider the scenario of $S_A=1$ and $S_B=3/2$. For~this case, the~eigenvalues and eigenstates of the Hamiltonian in Equation~(\ref{hamil}) is given in Table~\ref{tabone3} of   Appendix \ref{app1}. Again, we take the same measurement choices as before, i.e.,~measurement of $S_x$ spin operator on the side $A$ and $S_z$ spin operator on the side $B$.
We calculate $W$, $Q_M$ and the engine efficiency for each case. As in the previous asymmetric case, the~condition  $B_2>B_1$ does not guarantee the positivity of the extracted work $W_t$ in both   present cases. Starting~from a certain value of $J$, $W_t$ can be negative for $B_2>B_1$ and positive for $B_1>B_2$. Thus, for~asymmetric cases, $B_1>B_2$ and $B_2>B_1$, both   situations give rise to the extracted work to be negative staring from certain ranges of $J$.
We plot the efficiency of the heat engine for the aforesaid three asymmetric cases together in Figure~\ref{plot1sp13} for $B_2>B_1$ with $B_2=4$ and $B_1=3$. $B_1>B_2$ case can be calculated in a similar~way. 
%\subsection{Asymmetric scenario : $s_A=1$, $S_B=3/2$.}
We observe that, for asymmetric situation, the~efficiency becomes negative after a certain value of $J$. 
%\begin{figure}[h]
%\begin{center}
%\includegraphics[height=5cm,width=7cm]{neg3.pdf}
%\caption{(Color online) $-W_1$, $-W_2$, $W_t$ vs $J$ plot for $B_2>B_1$, with $B_2=4, B_1=3$. Here, $S_A=1$, $S_B=3/2$.}
%\label{negv3}
%\end{center}
%\end{figure}
%\begin{figure}[h]
%\begin{center}
%\includegraphics[height=5cm,width=7cm]{neg3.eps}
%\hfill
%\includegraphics[height=5cm,width=7.5cm]{neg3.eps}
%\hfill
%\includegraphics[height=5cm,width=7.5cm]{neg11.eps}
%\caption{Contribution of work strokes in negative efficiency for $B_1=3$ and $B_2=4$ for the first three plots. Last plot is for $B_1=4$ and $B_2=3$.}
%\label{spin1}
%\end{center}
%\end{figure}
In addition,  in~Figure~\ref{plot1sp13}, we can observe that, as the difference of spins increases between the two sides, the~efficiency goes to more negative value. Thus, from~these observations, it is clear that, if the two spins on both sides are not the same, then efficiency can be negative. { Another interesting feature to notice from the plot is that, within~the range of $J$ where efficiency is positive,  higher differences of the spin values give larger gain in efficiency over the uncoupled one.}
We have to take correct coupling strength $J$, to~have a higher but positive work output from these measurement-based coupled higher spin coupled heat engines. 
As in the case of $S_A=1/2$ and $S_B=1$, we also plot the work done in two adiabatic strokes for these two asymmetric cases in Figures~\ref{highspinwork1} and  ~\ref{highspinwork2} and note that the negative contribution in the extracted work is due to the second work~stroke.
%\begin{figure}[H]
%\begin{center}
%\includegraphics[height=4cm,width=6cm]{plot1sp3.eps}
%\caption{ Efficiency vs $J$ plot for $s_A=1/2$, $S_B=1/2$,  $s_A=1/2$, $S_B=1$ and $s_A=1/2$, $S_B=3/2$.}
%\label{plot1sp3}
%\end{center}
%\end{figure}
%\begin{figure}[H]
%\begin{center}
%\includegraphics[height=5cm,width=7.5cm]{plot1sp13-n.pdf}
%\caption{(Color online) Efficiency vs $J$ plot for ($S_A=1/2$, $S_B=1$), ($S_A=1/2$, $S_B=3/2$) and ($S_A=1$, $S_B=3/2$) ; $B_2=4$ and $B_1=3$.}
%\label{plot1sp13}
%\end{center}
%\end{figure}
\unskip
\unskip
\unskip
\unskip
%\begin{figure}[h]
%\begin{center}
%\includegraphics[height=5cm,width=7cm]{pos111.pdf}
%\caption{(Color online) $-W_1$, $-W_2$, $W_t$ vs $J$ plot for $B_2<B_1$, with $B_2=3, B_1=4$. Here $S_A=1/2$, $S_B=1$.}
%\label{negv11}
%\end{center}
%\end{figure}
\begin{figure}[h]
\centering
%\end{minipage}
%\begin{minipage}{0.32\linewidth}
\includegraphics[height=3.5cm,width=4.9cm]{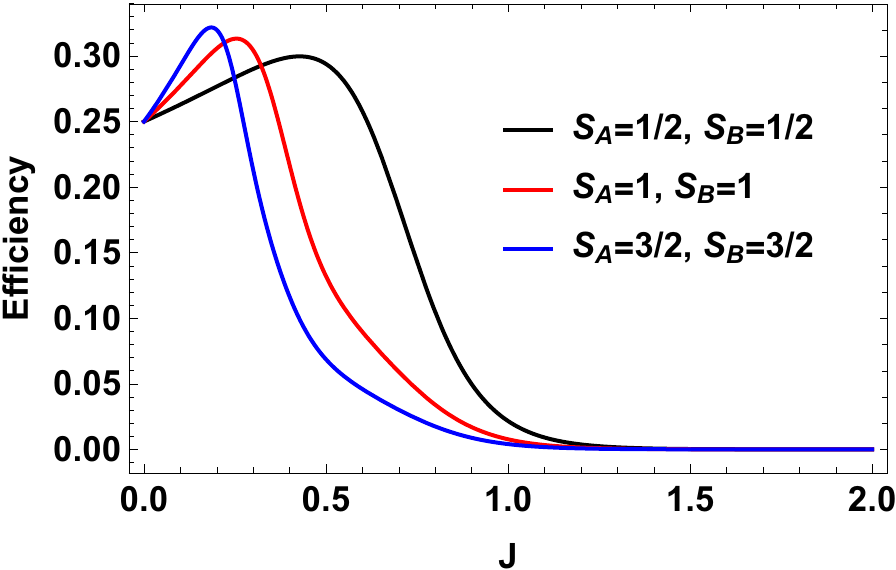}
\caption{(Color online) Efficiency vs. $J$ plot for ($S_A=1/2$, $S_B=1/2$), ($S_A=1$, $S_B=1$) and ($S_A=3/2$, $S_B=3/2$).}%This figure has been referred in section 5.2.
\label{plot1sp33}
%\end{minipage}
%\end{center}
\end{figure}

\subsection{Symmetric~Case} %$s_A=1$, $S_B=1$.}
In this section, we consider the symmetric case. One particular characteristic to look at is whether the efficiency gets negative for symmetric situation as well. We     show above one symmetric situation, namely  the~case of two spin-$1/2$ particles,   where we  see positive efficiency for all natures of $J$ (see Figure~\ref{spinhalf1}). Let us  investigate the case for higher spin symmetric situations. Three cases can arise for the symmetric scenario, if~we restrict ourselves up to spin-$3/2$. 
Among these, two spin-$1/2$ case is   discussed at the very beginning of the paper. The remaining two cases are the cases of two spin-$1$ and two spin-$3/2$ particles.
Eigenvalues and the eigenstates of the Hamiltonian Equation~(\ref{hamil}) for these two cases are given in     Tables~\ref{taboneone} and \ref{tab33}, respectively, in   Appendix \ref{app1}. We take the same measurement settings as before, i.e., measurement of $S_x$ spin operator on the side $A$ and $S_z$ spin operator on the side $B$.
We calculate $W$, $Q_M$, and~the engine efficiency for all these symmetric cases. 
Interestingly, for~the symmetric cases, for~$B_2>B_1$, we get the extracted work $W_t=-W$ to be positive for all $J$, and,~for $B_1>B_2$, $W_t$ is negative for all $J$. This is   the same  as in the case of two spin-$1/2$ particles (which is the simplest example of a symmetric case). For~two spin-$1$ particles, the~expression for $W_t$ is given by,
\begin{eqnarray}
\nonumber
&W_t=-W=p/q, \hspace{2mm} \text{where},\\
\nonumber
&p=(B2 - B1) (e^{4 B1}-1) (2 + e^{2 B1} + 2 e^{4 B1}\\
\nonumber
&+ e^{
   2 (B1 + 8 J)}),\\
\nonumber
\mbox{and,}\quad
&q=1 + e^{2 B1} + e^{4 B1} + e^{6 B1} + e^{8 B1} +\\
\nonumber
&e^{4 (B1 + 6 J)} + 
 e^{2 (B1 + 8 J)} (1 + e^{2 B1} + e^{4 B1}).
\end{eqnarray} 

From the above expression for $W_t$, it is evident that, whenever $B_2>B_1$, we have the extracted work  is positive and, for $B_1>B_2$, the~extracted work is negative. The situation for the other symmetric case is similar, i.e.,~for two spin-$3/2$ particles. One can also analyze the contributions of two adiabatic work strokes and find that sum of these two strokes gives rise to the extracted work a positive quantity for $B_2>B_1$.
We plot the efficiency of the engine for all symmetric cases together in Figure~\ref{plot1sp33} for $B_2>B_1$, with~$B_2=4$ and $B_1=3$. 
%\begin{figure}[H]
%\begin{center}
%\includegraphics[height=4cm,width=6cm]{plot1sp11.eps}
%\caption{ Efficiency vs $J$ plot for $s_A=1/2$, $S_B=1/2$ and $s_A=1$, $S_B=1$}
%\label{plot1sp13}
%\end{center}
%\end{figure}
%\subsection{Symmetric scenario : $s_A=1$, $S_B=1$.}
%\begin{figure}[H]
%\begin{center}
%\includegraphics[height=5cm,width=7.5cm]{plot1sp33-n.pdf}
%\caption{(Color online) Efficiency vs $J$ plot for ($S_A=1/2$, $S_B=1/2$), ($S_A=1$, $S_B=1$) and ($S_A=3/2$, $S_B=3/2$)}
%\label{plot1sp33}
%\end{center}
%\end{figure}
Given the fact that $Q_M$ is always positive, for~symmetric case, we~\textit{always} get positive efficiency for $B_2>B_1$. 
We also see that efficiency gets higher with the increase of spin value, although~the efficiency decreases more quickly  for higher spins with respect to the coupling strength~$J$.

Thus, from~these observations, we can conclude that, for the asymmetric scenario, the~work output is not always positive. However,~for the symmetric scenario, this is not the case. For~this latter scenario, we always get positive work output and hence positive efficiency for the heat engine with $B_2>B_1$. In addition, for higher spin scenario, the~efficiency is always greater (as long as it is positive for the asymmetric case) than that of the two spin half case. Thus, along with the coupling $J$, spin also plays an important role for the increase in efficiency for the measurement driven single temperature coupled heat~engine.
\section{Local vs. Global~Work}
\label{sec7}
In this section, we briefly touch upon the status of ``local'' and ``global'' works and how they are related. Until  now, we have been discussing the global aspect of the heat engine, i.e., global work output,  global efficiency, etc. Here, global means that we consider the coupled working medium as a whole and do not take into account the subsystems' contribution. In~the scenario of coupled quantum Otto cycle, the idea of local and global works was introduced by \citet{george} and further analyzed by \citet{ferdi}. During~the two thermalization strokes, states of the subsystems were defined by the reduced density matrices of the global thermal state of the coupled working medium. Local heat exchange is defined as the average energy transferred between the subsystem and the hot or cold heat bath  with respect to  the local Hamiltonian of the subsystem. According to the first law of thermodynamics, local work done by a subsystem is defined as the sum of the local heat exchange with the hot and cold heat bath. This definition is valid in the weak coupling approximation between the working medium and the heat bath. For~the measurement driven heat engine,  we define the local work in a similar  way. 
Let us  assume that, before the measurement, total state of the system is $\rho_{int}$ and after the measurement it becomes $\rho_{M}$. We denote  $\rho^A_{int}=Tr_B(\rho_{int})$ to be the reduced density matrix for the subsystem $A$ and $\rho^B_{int}=Tr_A(\rho_{int})$ to be the reduced density matrix of subsystem $B$, and~similarly for the reduced states after the measurement. In~the same way, we can calculate the reduced density matrix for the subsystems before and after the thermalization step. The~local work outputs for the subsystems are defined as~\cite{george, ferdi}, $w_i=-(q^i_1+q^i_2)$, where      
\begin{eqnarray}
&q^i_1=Tr[(\rho^i_{M}-\rho^i_{int})H^i(B_2)], \hspace{2mm} i=A,B;\\
&q^i_2=Tr[(\rho^i_{int}-\rho^i_{M})H^i(B_1)], \hspace{2mm} i=A,B.
\end{eqnarray}

\begin{figure}[h]
\centering
%\begin{center}
%\begin{minipage}{0.32\linewidth}
\includegraphics[height=3.5cm,width=4.4cm]{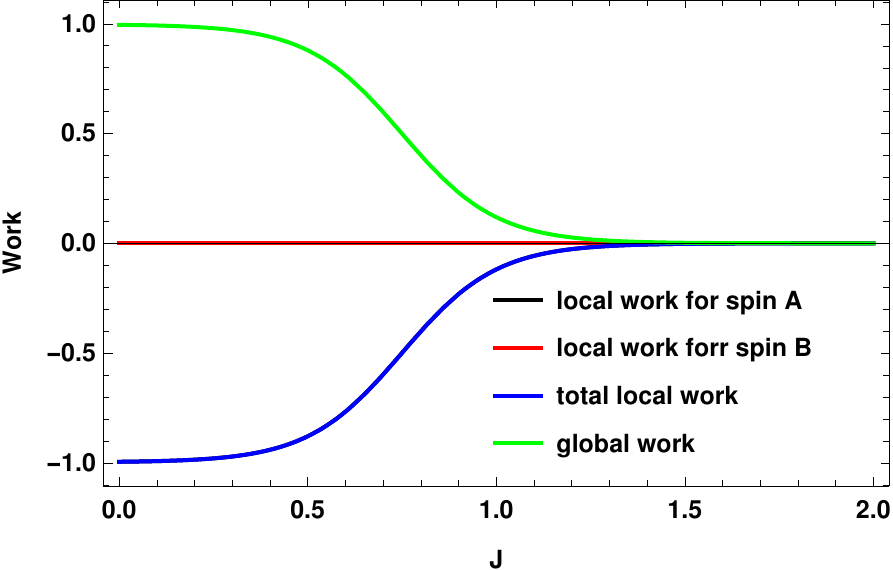}
\caption{(Color online) $B_2>B_1$, with~$B_2=4, B_1=3$. Here, $S_A=1/2$, $S_B=1/2$. { Blue and black curve have merged together.}}
\label{negv2}
\end{figure}

\begin{figure}[h]
\centering
%\end{minipage}
%\hfill
%\begin{minipage}{0.32\linewidth}
\includegraphics[height=3.5cm,width=4.4cm]{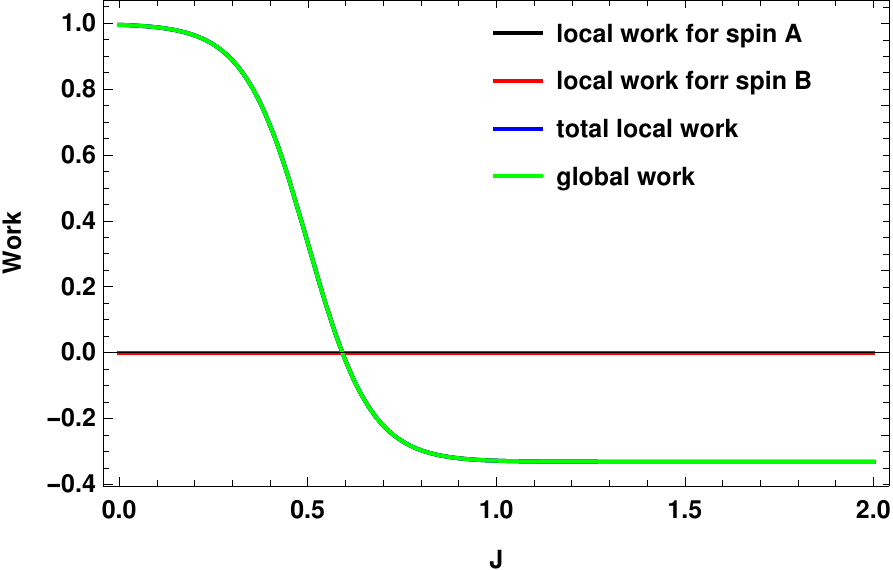}
\caption{(Color online) $B_2>B_1$, with~$B_2=4, B_1=3$. Here, $S_A=1/2$, $S_B=1$. { Black, blue and green curve have merged together.}}
\label{negv3}
\end{figure}

\begin{figure}[h]
\centering
%\end{minipage}
%\begin{minipage}{0.32\linewidth}
\includegraphics[height=3.5cm,width=4.4cm]{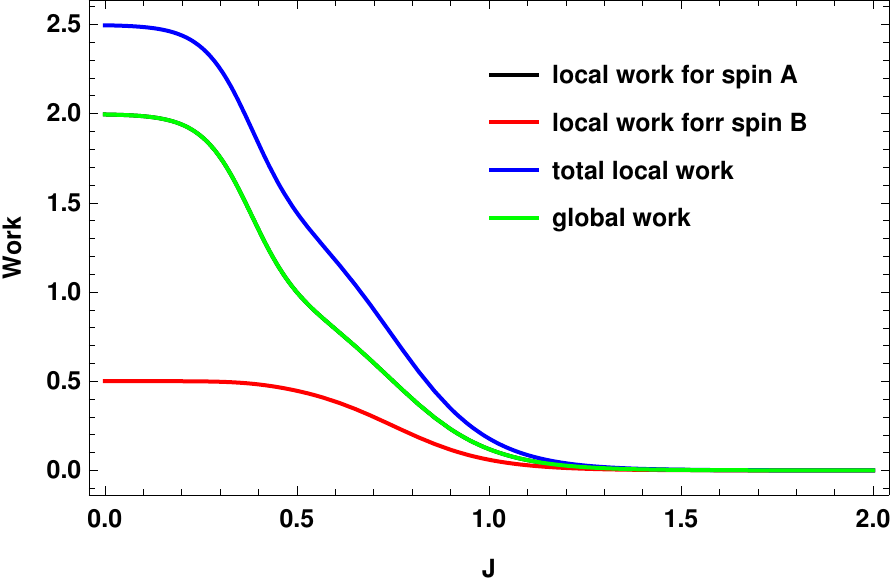}
\caption{(Color online) $B_2>B_1$, with~$B_2=4, B_1=3$. Here, $S_A=1$, $S_B=1$. { Black and green curve have merged together.}}%This figure has not been referred to within the text of the manuscript.----Now it has been referred. See the blue highlighted portion on page 16.
\label{negv4}
%\end{minipage}
%\end{center}
\end{figure}

$q^i_1$ represents the average energy exchange for the subsystem $A$ or $B$ ($i=A,B$) for the measurement step after the first adiabatic expansion and $q^i_2$ is the conventional heat exchange with the heat bath in the last step, i.e.,~the thermalization step. $H^i(B_1)$ and $H^i(B_2)$ are the local Hamiltonians for the subsystems ($i=A,B$) for external magnetic field $B_1$ and $B_2$, respectively. 
After the first adiabatic expansion, the~parameter of the Hamiltonian is changed from $B_1$ to $B_2$.  In~the next adiabatic stroke, the~magnetic field is changed back to the initial value $B_1$. Total local work done by the two subsystems is $w=w_A+w_B$.  Nevertheless, the~validity of these definitions in the scenario of measurement driven engine is under question as, during the measurement stroke, the~coupling between the working medium and the apparatus is in general not weak. Moreover, after the non-selective measurement, the~working medium is driven out of equilibrium, such that the state of the whole working medium is no longer a thermal state. 
In Figures~\ref{negv2}, \ref{negv3} and \ref{negv4}, we   plot  the local works and global works for three different spin combinations with same measurement settings considered above ($S_x$ on side $A$ and $S_z$ on side~$B$). The~plots show very interesting behavior. Unlike the coupled quantum  Otto cycle~\cite{george,ferdi}, the sum of the local works for the subsystems {is not always equal to the global work}. The~nature of the plots also change with the change of spin values. For~the two spin half case, {the sum of the local works start from a negative value and goes up to zero},  whereas, for the two other   scenarios, it starts with a positive value. {Interestingly, local work for one subsystem is zero for first two scenarios (Figures \ref{negv2} and \ref{negv3}}). Moreover, except~for the two spin half scenario, local work for one subsystem exactly matches with the global work (at least~for the cases considered here). { As a result, global work matches with the total local work for the second scenario (Figure \ref{negv3}) (black, blue and green curve have merged together)}. In~contrast to the conventional quantum  Otto cycle, the~relationship between local and global works (assuming~the above definition of local work) is complex. This also opens up the avenue for suitable definition of local work, taking into account the work cost for measurement, which demands a more detailed description of the engine cycle including the measuring apparatus. This may be a potential future area to~explore.
%%%%
\unskip
\unskip

\section{Conclusions}
\label{conclu}
In this paper, we investigate  the effect of coupled working medium in the measurement based single temperature quantum heat engine without feedback. We consider z one-dimensional Heisenberg model of two spins and calculate the efficiency of the heat engine. We show that, when the coupling constant $J$ is nonzero, the~efficiency is increased over the uncoupled scenario, i.e.,~$J=0$ case for certain range of non-zero $J$. Thus, interaction enhances the efficiency of this type of heat engine, similar to the coupled Quantum Otto cycle~\cite{george, ferdi}. We also consider the higher-dimensional scenario, where the two spins are not only spin-$1/2$ but also $1$ or $3/2$. In~these cases,  we observe a very interesting situation   that is  absent in the conventional coupled quantum Otto engine as well as in the case of uncoupled measurement driven heat engine. When the two spins of the two subsystems are the same, we always get useful work output from the engine, and~hence the efficiency   is positive. However,~this is not true for the asymmetric situation. If~the spins for the two subsystems are different, we~  get negative efficiency after a certain nonzero value of the coupling constant, implying that we cannot extract work, but~we have to invest work to run the cycle. It is very   similar to the case of a refrigerator in the absence of cold reservoir, which is replaced here by a measurement protocol. However,~as long as the efficiency is positive, for~higher spin systems, efficiency is higher than the lower-dimensional system. Thus, both coupling and dimension of the Hilbert space decide the efficiency of the measurement-based heat engine. Next, we consider  the local work and global work for the engine cycle. Using the same definition of local work used for quantum otto cycle, we show  that global work is never the same as total local work. This is unlike the results obtained in    ~\cite{george,ferdi}, which may be due to inappropriate definition of local work in the measurement driven heat engine.
Throughout the paper, we   focuss  on the quasistatic regime for the engine. In~    \cite{talkner2}, the~authors considered the scenario of imperfect thermalization stroke and analyzed the power of a single temperature measurement driven engine. Thus, the~effect of a coupled working medium on the power of this engine might be a good candidate for subsequent~study. 

\vspace{6pt}

%\authorcontributions{A.D formulated the work, carried out the calculations and prepared the manuscript. S.G supervised the work and checked the manuscript. Both authors discussed and examined the results.}%For manuscripts with more than one author, please state individual contributions of each author to research and writing of the manuscript.

%\funding{Interdisciplinary Cyber Physical Systems (ICPS) program  of the Department of Science and Technology (DST), India, Grant No.  DST/ICPS/QuEST/Theme-1/2019/13.}%Please disclose any funding information, or add "This research received no external funding."
\section{Acknowledgments}
The authors would like to thank G. De Chiara for bringing  to their notice the work reported in Ref.~\cite{chiara}. We acknowledge the support from Interdisciplinary Cyber Physical Systems (ICPS) program  of the Department of Science and Technology (DST), India, Grant No.  DST/ICPS/QuEST/Theme-1/2019/13.
%\section{Acknowledgments}

%\conflictsofinterest{The authors declare no conflicts of interest}%Please disclose conflicts of interest, or add “The authors declare no conflicts of interest.”

%\appendixtitles{yes}
%\appendix
\section*{Appendix}
\section{Tables\label{app1}}
\label{tables}
We write down the tables for both symmetric and asymmetric situations listing all the eigenvalues and eigenstates for the Hamiltonian in Equation~(\ref{hamil}). The first and third tables are for asymmetric cases, whereas the second and fourth tables are for symmetric~cases.
\begin{table}[h]
\centering
\caption{$S_A=1/2$, $S_B=3/2$.}\label{tabhalf3}
\begin{tabular}{cccc}
%\hline
%Acetylene: hydrogen concentration & \multicolumn{3}{|r|}{Average roughness values $S_a$ in nm} \\[0.5 ex]
%\hline
%{} &{|r|}{Value of $p$ to start violation} \\[0.3 ex]
\toprule
%{} & {}\\
%{} &\multicolumn{2}{|r|}{Value of $p$ to start violation} \\[0.3 ex]
%\hline
%\cline{1-2}
{\textbf{Eigenvalues}}  &   \textbf{Eigenstates} \\
%\midrule
\hline
$-2B-10J$ &$-\sqrt{\frac{3}{2}}\ket{0_A3_B}+\frac{1}{2}\ket{1_A2_B}=\ket{\psi_1}$\\

$2B-10J$ & $-\frac{1}{2}\ket{0_A1_B}+\sqrt{\frac{3}{2}}\ket{1_A0_B}=\ket{\psi_2}$\\

$-10J$ & $-\sqrt{\frac{1}{2}}\ket{0_A2_B}+\sqrt{\frac{1}{2}}\ket{1_A1_B}=\ket{\psi_2}$\\

$6J$ & $\sqrt{\frac{1}{2}}\ket{0_A2_B}+\sqrt{\frac{1}{2}}\ket{1_A1_B}=\ket{\psi_4}$\\

$-4B+6J$ & $\ket{1_A3_B}=\ket{\psi_5}$\\

$-2B+6J$ & $\frac{1}{2}\ket{0_A3_B}+\sqrt{\frac{3}{2}}\ket{1_A2_B}=\ket{\psi_6}$\\

$2B+6J$ & $\sqrt{\frac{3}{2}}\ket{0_A1_B}+\frac{1}{2}\ket{1_A0_B}=\ket{\psi_7}$\\

$4B+6J$ & $\ket{0_A0_B}=\ket{\psi_8}$\\
%\bottomrule
\hline
\end{tabular}
\end{table}
\unskip
\begin{table}[h]
\centering
\caption{$S_A=1$, $S_B=1$.}\label{taboneone}
\begin{tabular}{cccc}
\toprule
{\textbf{Eigenvalues}}  &   \textbf{Eigenstates} \\
%\midrule
\hline
$-2B-8J$ &$-\sqrt{\frac{1}{2}}\ket{1_A2_B}+\frac{1}{2}\ket{2_A1_B}$\\

$2B-8J$ & $-\sqrt{\frac{1}{2}}\ket{0_A1_B}+\sqrt{\frac{1}{2}}\ket{1_A0_B}$\\

$-4B+8J$ & $\ket{2_A2_B}$\\

$-16J$ & $\sqrt{\frac{1}{3}}\ket{0_A2_B}-\sqrt{\frac{1}{3}}\ket{1_A1_B}+\sqrt{\frac{1}{3}}\ket{2_A0_B}$\\

$-8J$ & $-\sqrt{\frac{1}{2}}\ket{0_A2_B}+\sqrt{\frac{1}{2}}\ket{2_A0_B}$\\

$8J$ & $\sqrt{\frac{1}{6}}\ket{0_A2_B}+\sqrt{\frac{2}{3}}\ket{1_A1_B}+\sqrt{\frac{1}{6}}\ket{2_A0_B}$\\

$4B+8J$ & $\ket{0_A0_B}$\\

$-2B+8J$ & $\sqrt{\frac{1}{2}}\ket{1_A2_B}+\sqrt{\frac{1}{2}}\ket{2_A1_B}$\\

$2B+8J$ & $\sqrt{\frac{1}{2}}\ket{0_A1_B}+\sqrt{\frac{1}{2}}\ket{1_A0_B}$\\
%\bottomrule
\hline
\end{tabular}
\end{table}
\unskip
\begin{table}[h]
\centering
\caption{$S_A=1$, $S_B=3/2$.}\label{tabone3}
\begin{tabular}{cccc}
\toprule
{\textbf{Eigenvalues}}  &   \textbf{Eigenstates} \\
%\midrule
\hline
$-B-20J$ &$\sqrt{\frac{1}{2}}\ket{0_A3_B}-\sqrt{\frac{1}{3}}\ket{1_A2_B}+\sqrt{\frac{1}{6}}\ket{2_A1_B}$\\

$B-20J$ & $\sqrt{\frac{1}{6}}\ket{0_A2_B}-\sqrt{\frac{1}{3}}\ket{1_A1_B}+\sqrt{\frac{1}{2}}\ket{2_A0_B}$\\

$-3B-8J$ & $-\sqrt{\frac{3}{5}}\ket{1_A3_B}+\sqrt{\frac{2}{5}}\ket{2_A2_B}$\\

$-B-8J$ & $-\sqrt{\frac{2}{5}}\ket{0_A3_B}-\sqrt{\frac{1}{15}}\ket{1_A2_B}+\sqrt{\frac{8}{15}}\ket{2_A1_B}$\\

$B-8J$ & $-\sqrt{\frac{8}{15}}\ket{0_A2_B}+\sqrt{\frac{1}{15}}\ket{1_A1_B}+\sqrt{\frac{2}{5}}\ket{2_A0_B}$\\

$3B-8J$ & $-\frac{2}{5}\ket{0_A1_B}+\sqrt{\frac{3}{5}}\ket{1_A0_B}$\\

$-3B+12J$ & $\sqrt{\frac{2}{5}}\ket{1_A3_B}+\frac{3}{5}\ket{2_A2_B}$\\

$3B+12J$ & $\sqrt{\frac{3}{5}}\ket{0_A1_B}+\sqrt{\frac{2}{5}}\ket{1_A0_B}$\\

$-5B+12J$ & $\ket{2_A3_B}$\\

$-B+12J$ & $\sqrt{\frac{1}{10}}\ket{0_A3_B}+\sqrt{\frac{3}{5}}\ket{1_A2_B}+\sqrt{\frac{3}{10}}\ket{2_A1_B}$\\

$B+12J$ & $\sqrt{\frac{3}{10}}\ket{0_A2_B}+\sqrt{\frac{3}{5}}\ket{1_A1_B}+\sqrt{\frac{1}{10}}\ket{2_A1_B}$\\

$5B+12J$ & $\ket{0_A0_B}$\\
%\bottomrule
\hline
\end{tabular}
\end{table}
\unskip
\begin{table}[H]
%\begin{center}
\centering
\caption{$S_A=3/2$, $S_B=3/2$.}\label{tab33}
\begin{tabular}{cccc}
%\hline
%Acetylene: hydrogen concentration & \multicolumn{3}{|r|}{Average roughness values $S_a$ in nm} \\[0.5 ex]
%\hline
%{} &{|r|}{Value of $p$ to start violation} \\[0.3 ex]
\toprule
%{} & {}\\
%{} &\multicolumn{2}{|r|}{Value of $p$ to start violation} \\[0.3 ex]
%\hline
%\cline{1-2}
\textbf{Eigenvalues}  &   \textbf{Eigenstates}\\
%\midrule
\hline
$-2B-22J$ &$\sqrt{\frac{3}{10}}\ket{1_A3_B}-\sqrt{\frac{2}{5}}\ket{2_A2_B}+\sqrt{\frac{3}{10}}\ket{3_A1_B}$\\

$2B-22J$ & $\sqrt{\frac{3}{10}}\ket{0_A2_B}-\sqrt{\frac{2}{5}}\ket{1_A1_B}+\sqrt{\frac{3}{10}}\ket{2_A0_B}$\\

$-4B-6J$ & $-\sqrt{\frac{1}{2}}\ket{2_A3_B}+\sqrt{\frac{1}{2}}\ket{3_A2_B}$\\

$-2B-6J$ & $-\sqrt{\frac{1}{2}}\ket{1_A3_B}+\sqrt{\frac{1}{2}}\ket{3_A1_B}$\\

$-6B+18J$ & $\ket{3_A3_B}$\\

$2B-6J$ & $-\sqrt{\frac{1}{2}}\ket{0_A2_B}+\sqrt{\frac{1}{2}}\ket{2_A0_B}$\\

$4B-6J$ & $-\sqrt{\frac{1}{2}}\ket{0_A1_B}+\sqrt{\frac{1}{2}}\ket{1_A0_B}$\\

$-30J$ & $-\frac{1}{2}\ket{0_A3_B}+\frac{1}{2}\ket{1_A2_B}-\frac{1}{2}\ket{2_A1_B}+\frac{1}{2}\ket{3_A0_B}$\\

$-22J$ & $\frac{3}{\sqrt{20}}\ket{0_A3_B}-\frac{1}{\sqrt{20}}\ket{1_A2_B} -\frac{1}{\sqrt{20}}\ket{2_A1_B}+\frac{3}{\sqrt{20}}\ket{3_A0_B}$\\

$-6J$ & $-\frac{1}{2}\ket{0_A3_B}-\frac{1}{2}\ket{1_A2_B}+\frac{1}{2}\ket{2_A1_B}+\frac{1}{2}\ket{3_A0_B}$\\

$18J$ & $\frac{3}{\sqrt{20}}\ket{0_A3_B}+\frac{1}{\sqrt{20}}\ket{1_A2_B}+\frac{1}{\sqrt{20}}\ket{2_A1_B}+\frac{3}{\sqrt{20}}\ket{3_A0_B}$\\

$6B+18J$ & $\ket{0_A0_B}$\\

$-4B+18J$ & $\sqrt{\frac{1}{2}}\ket{2_A3_B}+\sqrt{\frac{1}{2}}\ket{3_A2_B}$\\

$-2B+18J$ & $\sqrt{\frac{1}{5}}\ket{1_A3_B}+\sqrt{\frac{3}{5}}\ket{2_A2_B}+\sqrt{\frac{1}{5}}\ket{3_A1_B}$\\

$2B+18J$ & $\sqrt{\frac{1}{5}}\ket{0_A2_B}+\sqrt{\frac{3}{5}}\ket{1_A1_B}+\sqrt{\frac{1}{5}}\ket{2_A0_B}$\\

$4B+18J$ & $\sqrt{\frac{1}{2}}\ket{0_A1_B}+\sqrt{\frac{1}{2}}\ket{1_A0_B}$\\
%\bottomrule
\hline
%\end{center}
\end{tabular}
\end{table}

%\section{Symmetric cases}

%\columnbreak

%$\ket{0_{A/B}}\doteq
%\begin{pmatrix} 1\\0\end{pmatrix}$ and $\ket{1_{A/B}}\doteq
%\begin{pmatrix} 0\\1\end{pmatrix}$ are the eigenstates of $S_{A/B}^z$ for spin-$1/2$ particle. $\ket{0_{A/B}}\doteq
%\begin{pmatrix} 1\\0\\0\end{pmatrix}$, $\ket{1_{A/B}}\doteq
%\begin{pmatrix} 0\\1\\0\end{pmatrix}$ and $\ket{2_{A/B}}\doteq
%\begin{pmatrix} 0\\0\\1\end{pmatrix}$ are the eigenstates of $S_{A/B}^z$ for spin-$1$ particle. $\ket{0_{A/B}}\doteq
%\begin{pmatrix} 1\\0\\0\\0\end{pmatrix}$, $\ket{1_{A/B}}\doteq
%\begin{pmatrix} 0\\1\\0\\0\end{pmatrix}$, $\ket{2_{A/B}}\doteq
%\begin{pmatrix} 0\\0\\1\\0\end{pmatrix}$ and $\ket{3_{A/B}}\doteq
%\begin{pmatrix} 0\\0\\0\\1\end{pmatrix}$ are the eigenstates of the operator $S_{A/B}^z$ for spin-$3/2$ particle.

%%%%%%%%%%%%%%%%%%%%%%%%%%%%%%%%%%%%%%%%%%%%%%%%%%%

%\section*{References}
%\centering

\end{document}